\documentclass[twocolumn,showpacs,preprintnumber,amsmath,amssymb,nofootinbib]{revtex4}

\usepackage{epsfig}
\usepackage{graphicx, subfigure}
\usepackage{epstopdf}

\usepackage[colorlinks]{hyperref}

\usepackage{graphicx}

\usepackage{bm}
\usepackage[normalem]{ulem}
\usepackage[inline]{enumitem}
\usepackage{amsfonts}
\usepackage{amsmath}
\usepackage{amssymb}

\newcommand{\be}{\begin{equation}}
\newcommand{\ee}{\end{equation}}

\newcommand\aap{Astron. Astrophys. }
\newcommand\jcap{J. Cosmol. Astropart. Phys. }
\newcommand\mnras{Mon. Not. R. Astron. Soc. }
\newcommand\apjs{Astrophys. J. Suppl. Ser. }
\newcommand\grg{Gen. Relativ. Gravit. }
\newcommand\etal{{\it et al.}}

\graphicspath{{./}}

\newcommand{\Mov}[1]{{\color{red}{#1}}}

\begin{document}
\title{Dynamical friction shear and rotation in Chaplygin cosmology}

\author{A. Del Popolo}
\email{adelpopolo@astro.iag.usp.br}
\affiliation{Dipartimento di Fisica e Astronomia, Universit\'a di Catania, Italia}
\affiliation{
INFN sezione di Catania,
	Via S. Sofia 64, I-95123 Catania, Italy}
\author{Saeed Fakhry}
\email{s\_fakhry@kntu.ac.ir}
\affiliation{Department of Physics, K.N. Toosi University of Technology, P.O. Box 15875-4416, Tehran, Iran}

\author{Maryam Shiravand}
\email{ma\_shiravand@kntu.ac.ir}
\affiliation{Department of Physics, K.N. Toosi University of Technology, P.O. Box 15875-4416, Tehran, Iran}
\author{Morgan Le Delliou}%
\email{delliou@lzu.edu.cn,\\Corresponding author, Morgan.LeDelliou.IFT@gmail.com}
\affiliation{%
Institute of Theoretical Physics \& Research Center of Gravitation, Lanzhou University, Lanzhou 730000, China
}
\affiliation{
Key Laboratory of Quantum Theory and Applications of MoE, Lanzhou University, Lanzhou 730000, China
}
\affiliation{
Lanzhou Center for Theoretical Physics \& Key Laboratory of Theoretical Physics of Gansu Province, Lanzhou University, Lanzhou 730000, China
}
\affiliation{%
Instituto de Astrof\'isica e Ci\^encias do Espa\c co, Universidade de Lisboa, Faculdade de Ci\^encias, Ed.~C8, Campo Grande, 1769-016 Lisboa, Portugal
}
\affiliation{%
Universit\'e de Paris-Cit\'e, APC-Astroparticule et Cosmologie (UMR-CNRS 7164), 
 F-75006 Paris, France}

\date{\footnotesize{Received \today; accepted ?}}

\date{\today}
\begin{abstract}
In this study, we build upon the findings of Del Popolo et al. (2013) by further analyzing the influence of dynamical friction on the evolution of cosmological perturbations within the framework of the spherical collapse model (SCM) in a Universe dominated by generalized Chaplygin gas (GCG). Specifically, we investigate how dynamical friction alters the growth rate of density perturbations, the effective sound speed, the equation-of-state parameter www, and the evolution of the cosmic expansion rate. Our results demonstrate that dynamical friction significantly delays the collapse process compared to the standard SCM. Accurate computation of these parameters is crucial for obtaining consistent results and reliable physical interpretations when employing the GCG model. Furthermore, our analysis confirms that the suppression of perturbation growth due to dynamical friction is considerably more pronounced than that caused by shear and rotation, as previously indicated by Del Popolo et al. (2013). This enhanced suppression effectively addresses the instability issues, such as oscillations or exponential divergences in the dark-matter power spectrum, highlighted in linear perturbation studies, such as those by Sandvik et al. (2004).
\end{abstract}


\maketitle

\section{Introduction} \label{sec:Introduction} 

Nowadays, the standard model of cosmology, also dubbed $\Lambda$ cold dark matter ($\Lambda$CDM) model gives predictions
in a very good agreement with observations expecially on cosmological, and intermediate scales \citep{Komatsu2011,Spergel,Kowalski,Percival}. However the model suffers from some issues. At large scales several tensions are present. Here is a non exhaustive list: \begin{enumerate*}[label=\itshape\alph*.]
                                                                                                                                                                                                                                                                                                              \item the Hubble tension: the discrepancy in value of the Hubble parameter, $H_0$, as measured by the early time probes (e.g., CMB), and late time, model-independent, determination of $H_0$ obtained from local measurements of distances and redshifts \citep{DiValentino}; 
                                                                                                                                                                                                                                                                                                              \item the tension between the Planck 2015 data with $\sigma_8$ growth rate \citep{Macaulay}, and with CFHTLenS weak lensing \citep{Raveri} data; 
                                                                                                                                                                                                                                                                                                              \item the small scales (1-10 kpcs) issue \citep{DelPopolo2017}.

\end{enumerate*}
 At cosmological scales the model suffers from the {\it cosmological constant problem}
 \citep{Weinberg,Astashenok}, and {\it the cosmic coincidence problem} \citep{Steinhardt}. The $\Lambda$CDM model assumes that the Universe is dominated by dark matter (DM) and an exotic component with a negative pressure, usually named dark energy (DE). This last component is responsible of the accelerated rate of expansion of the Universe, result obtained
from the observations of high redshift supernovae, dimmer than expectations \cite{SN}, and later confirmed by
independent observations like the angular spectrum of the CMBR temperature fluctuations \cite{CM} and the baryon
acoustic oscillations \cite{Teg}. After more than two decades the nature of dark matter is not known. Many models
have been published, and in the simplest DE is identified with the cosmological constant $\Lambda$, and vacuum energy. As reported, when $\Lambda$ is interpreted as the vacuum energy, a huge discrepancy, dubbed cosmological
constant problem, a factor $10^{123}$, is found between the vacuum energy obtained from quantum theory and
that obtained from observations. Because of the quoted problem, and the  cosmic coincidence problem, several other alternative DE models have been proposed: \begin{enumerate*}[label=\itshape\alph*.]
                                                                                                                                                                        \item the quintessence models, in which DE is related to a scalar field weakly interacting with matter \cite{Gum}; 
                                                                                                                                                                        \item K-essence, phantom models, or unified dark models (UDM) \cite[e.g.,][]{AvBe}. 
                                                                                                                                                                       \end{enumerate*}
 The interesting aspect of UDMs is that the main components of the Universe, DM and DE, are described by the same physical entity. A very interesting aspect of the Chaplygin gas (CG) is that it behaves as a dust-like matter (pressureless DM) at early times (high densities) and behaves like a cosmological constant at late times (low densities). As discussed in the Appendix, at early time ($a \rightarrow 0$), the CG behaves as DM, and at later times, as $w \rightarrow -1$, it approaches a DE behavior. Unfortunately, the quoted model has been under strong observational pressure from CMB anisotropies \citep{Sandvik,Bean}. This has led to the study of several interacting Chapligyn gas models in the literature \citep{Zhang,Saha}.
Kamenshchik \Mov{\it et al. }\cite{Kam}, introduced a peculiar case of UDMs, the generalized Chaplygin gas (GCG). Developed in several papers
\cite[e.g.][]{BBG,Bento2002a,Bento2004}. Several theoretical studies, cosmological tests using the X-ray luminosity of
galaxy clusters, CMB measurements, and lensing statistics have been performed \cite{Bord,Ben,Dev,Cun,Fab}. 
This GCG model has been confronted, finding consistency, with different tests involving type Ia supernovae (SNIa), cosmic microwave background (CMB), and other observational datasets \citep{Bento2003a,Bento2003b,Bertolami2004,Alcaniz2003,Bento2005,Barreiro2008}.
In the peculiar case of UDMs having $\alpha \neq 0$ (see appendix) comparison of the linear theory with observations have put in evidence some problems of the GCG UDM \cite{Mult,avelino:2004}. In GCG UDM non-linear effects generate a non trivial backreaction in the background dynamics, visible when studying the onset of the nonlinear regime in GCG UDMs \cite{avelino:2004} . As a consequence, for all $\alpha \neq 0$ models, is observed a break down of the linear theory at late times. Despite the quoted problem, there is a certain agreement between GCG UDM and large scale structure observations \cite{Bec}. If we want
to know if the GCG can be an alternative to the $\Lambda$CDM, we need to study the non-linear evolution of DM and
DE in the Chaplygin gas cosmology \cite{bilic:2004}. Apart from the fully non-linear analysis as performed in SPH simulations  \cite[see e.g.][]{maccio:2004,aghanim:2009,baldi:2010,li:2011}, the non-linear evolution of perturbations of DM and DE can be performed through the spherical collapse model \cite[SCM,][]{GG72,Fillmore1984,Bertschinger1985,Hoffman1985,Ryden1987,Avila-Reese1998,Subramanian2000,Ascasibar2004,Williams2004}. While the seminal paper of \citep{GG72} presented the SCM considering only radial collapse, several other authors showed how angular momentum can be included in the model \citep{Ryden1987,Gurevich1988a,
Gurevich1988b,White1992,Sikivie1997,Avila-Reese1998,Nusser2001,Hiotelis2002,LeDelliou2003,Ascasibar2004,Williams2004,
Zukin2010}. The SCM was used by Fernandes \Mov{\it et al. }\cite{Fern}  to perform the quoted non-linear analysis. Differently from other
works \cite[e.g.][]{bilic:2004, multamaki:2004, pace:2010} Fernandes \Mov{\it  et al.'s} treatment \cite{Fern}  considers the collapse of both GCG and baryons, in the post-recombination epoch (neglecting radiation), assuming a time-dependent equation-of-state parameter $w$, for the background and the collapsing region. Their study had a crucial limit, as rotation (vorticity), $\omega$, and shear, $\sigma$ were put equal to zero. In any proper extension of the SCM the contraction effect produced by shear and the expansion one produced by vorticity should be considered, as done by \cite{EN2000}. The previous authors studied the effect of
shear and vorticity in DM-only dominated universes, and only in \cite{DelPopolo2013}, shear and vorticity effects were considered in the case of DM and DE dominated universes. \citep{Fern}, studied the spherical $top-hat$ collapse framework in
GCG dominated universes, while in \citep{DelPopolo2013}, we extended the \citep{Fern} model taking into account shear and vorticity in the collapse. We showed that the collapse is slowed down in \cite{DelPopolo2013} for all quantities studied by \citep{Fern}. As we showed in several other papers \cite{Pace2025,DelPopolo:2006au}, if we want to have a more realistic SCM, we must take account of dynamical friction. In
the present paper, we extend \citep{DelPopolo2013} to take account of dynamical friction, and see how the collapse is modified.

The paper is organized as follows: Sec.~\ref{sec:SC} summarizes the model used. It reviews the derivation of the
equation of the SCM in presence of shear and vorticity, the {\em effective sound speed} used, and the way equations were
integrated. Sec.~\ref{sec:res} deals with results and Sec.~\ref{sec:conc} with conclusions.

\section{Model} 
\label{sec:SC}

In the seminal paper of \citep{GG72}, the authors studied the infall of matter into clusters of galaxies. Their treatment
supposed that the structure collapsed radially and discarded non-radial motions. Several following papers showed how to introduce non-radial motions, and angular momentum, $L$,
\citep{Ryden1987,Gurevich1988a,Gurevich1988b,White1992,Sikivie1997,Avila-Reese1998,Nusser2001,Hiotelis2002,  
LeDelliou2003,Ascasibar2004,Williams2004,Zukin2010} preserving spherical symmetry\footnote{Spherical symmetry is 
preserved if one assumes that the distribution of angular momenta of particles is random, so as to produce a net null mean
angular momentum \cite{White1992}.}. The equations of the SCM with angular momentum can be written as 
\citep[e.g.,][]{Peebles1993,Nusser2001,Zukin2010}:
\begin{equation}
\frac{d^2 R}{d t^2}= -\frac{GM}{R^2} +\frac{L^2}{M^2 R^3}.
\label{eqn:spher}
\end{equation}

The SCM was further extended in \citep{DelPopolo1998,DelPopolo2006,DelPopolo09,Lahav1991, Bartlett1993,Antonuccio1994,DelPopolo2019}, taking into account the cosmological constant, a particular form of dark energy, and dynamical friction:
\begin{equation}\label{eq:coll}
 \ddot{R} = -\frac{GM}{R^2} + \frac{L^2(R)}{M^{2}R^3} + \frac{\Lambda}{3}R -
 \eta\frac{{\rm d}R}{{\rm d}t}\,,
\end{equation}
being $\eta$ the dynamical friction coefficient. 

Adding the effects of shear, vorticity, and generalising DE, the previous equation can be written as
\begin{equation} \label{eqn:wnldeqqq}
\ddot{R} = -\frac{GM_{\rm m}}{R^2} - \frac{GM_{\rm de}}{R^2}(1+3w_{\rm de})-\frac{\sigma^2-\omega^2}{3}R -\eta \frac{dR}{dt}.
\end{equation}
being $M_{\rm m}= \frac{4 \pi R^3}{3} (\bar{\rho}+\delta \rho)$, $M_{\rm de}$ the mass of the dark-energy component enclosed in the volume, $\bar{\rho}_{\rm de}$, and $w_{\rm de}$ being respectively its background density and equation-of-state \citep{Fosalba1998,EN2000,Ohta2003}. $M_{\rm m}$, as shown, contains background and perturbation.
A similar equation (excluding the dynamical friction term) was obtained by several authors 
\citep[e.g.,][]{Fosalba1998,EN2000,DelPopolo2013a}.

By means of the relation $\delta=\frac{2GM_{\rm m}}{\Omega_{m,0} H^2_0}(a/R)^3-1$, Eq. (\ref{eq:coll}) can be written in terms
of the overdensity $\delta$. However, we choose to obtain the equation of evolution of $\delta$ using the Pseudo-Newtonian (PN) approach to cosmology \citep{jsal}.

\subsection{PN equations}

The evolution equations of $\delta$ in the non-linear regime, in the PN approach, has been obtained and used in the
framework of structure formation, spherical and ellipsoidal collapse, by several authors \citep{Bernardeau1994,Ohta2003,Ohta2004,Abramo2007}.
At this stage, we may recall that here we want to generalize the work of \citep{DelPopolo2013} taking into account dynamical
friction, and that in \citep{DelPopolo2013} we generalized the work of Fernandes \Mov{\it et al. }\citep{Fern} including the contributions from the shear and rotation terms. We therefore closely follow \citep{Fern,DelPopolo2013} to derive the needed equations, modified to take into account the effect of dynamical friction. We assume that the velocity of light is $c = 1$, and that the fluid satisfies the equation-of-state $P = w \rho$. The generalizations of the continuity
equation, of Euler's equation (both valid for each fluid species labelled $j$), and of Poisson's equation (which is valid
for the sum of all fluids) given by \citep{jsal,Abramo2007} are used, and expressed in terms of the density $\rho_j$, pressure $p_j$, velocity $\overrightarrow{u_j}$, and potential $\Phi$ \citep[see Eqs. 11-14 in][]{Abramo2007}.
In several paper, has been shown that dynamical friction manifests itself during the evolution of perturbations
and affects their evolution, slowing their collapse. As a consequence cosmic structures require more time to form.
As in Newtonian dynamics, friction affects the equations of motion and, being proportional to velocity, it modifies
the Euler equation, while not modifying the other equations \citep[see][]{Pace2025}.

Introducing cosmological perturbations in the previous equations, translating to comoving coordinates, $\vec{x}=\vec{r}/a$,
while defining $\delta_{\rm j}=\delta\rho_{\rm j}/\rho_{\rm j}$, and assuming that $w_{\rm j}$ and $c_{\rm eff,j}^{2}$ are
functions of time only, the equations for the perturbed quantities are:
\begin{eqnarray}
\dot{\delta}_{\rm j}+3H\left(c_{\rm eff,j}^{2}-w_{\rm j}\right)\delta_{\rm j} &=& \nonumber\\
-\left[1+w_{\rm j}+\left(1+c_{\rm eff,j}^{2}\right)\delta_{\rm j}\right]
\frac{\vec{\nabla}\cdot\vec{v}_{\rm j}}{a}-\frac{\vec{v}_{\rm j}\cdot\vec{\nabla}\delta_{\rm j}}{a},
\label{cont-pert2}
\end{eqnarray}
\begin{equation}
\dot{\vec{v}}_{\rm j}+(H+\eta)\vec{v}_{\rm j}+\frac{\vec{v}_{\rm j}\cdot\vec{\nabla}}{a}\vec{v}_{\rm j}
=-\frac{\vec{\nabla}\phi}{a}-\frac{c_{\rm eff,j}^{2}\vec{\nabla}\delta}
{a\left[1+w_{\rm j}+(1+c_{\rm eff,j}^{2}) \delta_{\rm j}\right]}\;,
\label{euler-pert2}
\end{equation}
\begin{equation}
\frac{\nabla^{2}\phi}{a^{2}}=4\pi G\sum_{\rm k}\rho_{0_{\rm k}}\delta_{\rm k}\left(1+3c_{\rm eff,k}^{2}\right)\;,
\label{poisson-pert2}
\end{equation}
where $\eta$ is the coefficient of dynamical friction, and $c^2_{\rm eff,j}\equiv\delta p_{\rm j}/\delta\rho_{\rm j}$ is the  effective sound speed of each fluid.

The previous equations can be simplified as in \cite{Abramo2007}:
\begin{eqnarray}
\dot\delta_{\rm j} & = & -3H(c^2_{\rm eff,j}-w_{\rm j})\delta_{\rm j} \nonumber\\
& & -[1+w_{\rm j}+(1+c^2_{\rm eff,j})\delta_{\rm j}]\frac{\theta_{\rm j}}{a}\;,
\label{eq:dot_delta1}\\
\dot\theta_{\rm j} & = & -H\theta_{\rm j}-\frac{\theta_{\rm j}^2}{3a} \nonumber\\
& & -4\pi Ga\sum\limits_{\rm k}{\rho_{0\rm k}\delta_{\rm k}(1+3c^2_{\rm eff,k})} \nonumber\\ 
& & -\frac{\sigma_{\rm j}^2-\omega_{\rm j}^2}{a}\;.
\label{eq:dot_theta1}
\end{eqnarray}
where $\theta_j \equiv \nabla \cdot \vec{v}_j$ and ${\vec v}_j$ is the peculiar velocity field.

The number of equations is equal to the number of cosmological fluid components in the
system. Shear and vorticity are already present in Eq.~\eqref{euler-pert2}, via the term
$(\vec{v}\cdot\vec{\nabla})\vec{v}$. To obtain Eq.~\eqref{eq:dot_theta1}, and the scalars $\sigma$ and $\omega$, one simply need to take the divergence of
Eq.~\eqref{euler-pert2}.

Recalling that the density parameters follow $\Omega_{\rm j}= \frac{8\pi G}{3H^2}\rho_{\rm 0j}$, the previous equations, in terms of the scale factor $a$, can be expressed in the form: 
\begin{eqnarray}
\delta_{\rm j}^{\prime} & = & -\frac{3}{a}(c^2_{\rm eff,j}-w_{\rm j})\delta_{\rm j}\nonumber\\
& & -[1+w_{\rm j}+(1+c^2_{\rm eff,j})\delta_{\rm j}]\frac{\theta}{a^2H}\;,
\label{eq:dot_delta_a}\\
\theta^{\prime} & = & -\frac{\theta}{a}-\frac{\eta \theta}{H}-\frac{\theta^2}{3a^2H}\nonumber\\
& & -\frac{3H}{2}\sum\limits_{\rm j}{\Omega_{\rm j}\delta_{\rm j}(1+3c^2_{\rm eff,j})}
- \frac{\sigma^2-\omega^2}{a^2 H}\;,
\label{eq:dot_theta_a_omega}
\end{eqnarray}
where the prime denotes the derivative with respect to $a$.

The way to evaluate the term $\sigma^2-\omega^2$ was discussed in \cite{DelPopolo2013a,DelPopolo2013,DelPopolo2020}
by defining the ratio $\beta$ between the rotational and gravitational term in Eq. (\ref{eqn:spher}):
\begin{equation}
\beta=\frac{L^2}{M^3 RG}\;.
\label{eqn:beta}
\end{equation}

Its values increases from galaxy clusters having $\beta \simeq 10^{-3}$ \citep{Mehrabi2017}, to dwarf galaxies size perturbations. For the Milky Way $\beta \simeq 0.4$. In order to obey to a value for $\beta$ similar to the one obtained by \citep{ST01}, we set $\beta= 0.04$ for galactic masses \citep[see also][]{DelPopolo2013a}.

It is possible to relate $\sigma^2-\omega^2$ to $\delta$, as follows \citep{DelPopolo2013,DelPopolo2013a,DelPopolo2020}:
\begin{equation}
\frac{\sigma^2-\omega^2}{a^2 H^2}=-\frac{3}{2}\beta\sum\limits_{\rm j}{\Omega_{\rm j}\delta_{\rm j}(1+3c^2_{\rm eff,j})}\;.
\label{eqn:so}
\end{equation}

Eq.~\eqref{eqn:beta} is based on the assumption that the ratio of acceleration due to the shear/rotation term to that of
the gravitational field, is constant during the collapse
. One could think that, since $L$, generated by
tidal torques, could decrease in the collapsing phase, producing a reduction of the value of $\beta$, this could undermine
the calculation. As discussed in \citep{DelPopolo2013}, angular momentum acquisition is maximum at turn-around, and later remains constant, since it is found not to be lost in the collapse phase. Moreover by the definition of $\beta$, since $M$ remains constant, and $R$ decrease in the collapse to a minimum of $R_{\rm final} \simeq R_{\rm initial}$,
$\beta$ increases in the collapse. Similarly to \citep{DelPopolo2013,DelPopolo2013a}, we will consider the cases $\beta = 0.04$ (Milky Way), and $\beta = 0.02$, and $\beta = 0.01$ (slower rotation).
In \citep{DelPopolo2013}, we solved a system of two fluids, modeled with Eqs.~\eqref{eq:dot_delta_a} (one for the
GCG and one for baryons), and with Eq.~\eqref{eq:dot_theta_a_omega} for $\bar{C}=0.75$ (see the Appendix, and the Table in \citep{Fern}), for $\alpha = 0$, model equivalent to the $\Lambda$CDM, $\alpha = 0.5$, and 1. As we soon find, $\eta = \eta_0 H$, and we use two values for $\eta_0$. The initial conditions (ICs) for the system, the values of the density parameters, and Hubble constant are the same as \citep{Fern}, and in agreement with values for the $\Lambda$CDM \cite{CM}.
As {in} \cite{Fern}, $p_{\rm b}=w_{\rm b}=c_{\rm s,b}^2=c_{\rm eff,b}^2=0$.

\subsection{Dynamical friction}

In this subsection, we show how dynamical friction evolves, and we obtain its typical values. For this we follow \citep{Antonuccio1994,DelPopolo09}, to which we refer for more details. Given a primordial Gaussian density field, in a scenario with hierarchical structure formation, structures of size $R$ form around the local maxima of the field smoothed over a scale of size $R$ \citep{Bardeen,Bond,Colafrancesco}. In such a system, the gravitational field to which a test particle is subject, can be decomposed in a part associated with the smoothed global mass distribution and a stochastic part which originates
from the particle number fluctuations and gives rise to a frictional force $-\eta {\bf v}$ where $\eta$ is the dynamical friction coefficient and ${\bf v}$ the macroscopic velocity. By means of the virial theorem, we may show that the dynamical friction coefficient is given by:
\begin{align}
 \eta = &\, \frac{4.44(Gm_{\rm a}n_{\rm a})^{1/2}}{N}\log{[1.12N^{2/3}]} \nonumber \\
 = &\, 4.44\sqrt{\frac{3\Delta}{8\pi}}\frac{\log{[1.12N^{2/3}]}}{N}H = \eta_0\,H\,,
\end{align}
where we used the relation $m_{\rm a}n_{\rm a}=\rho_{\rm m}=\bar{\rho}_{\rm m}\Delta$, with $\Delta$ the average overdensity of the perturbation. In principle, $\Delta$ would depend on the virialization recipe and on the cosmological model. However, here for simplicity, we will assume that it is constant, of the order of 100. 

We may calculate the value of $\eta_0$ knowing the total number of peaks $N$. If we consider a cluster, $N$ is considered to
be approximately constant \citep{Antonuccio1994}. For $N=10^3$, we find $\eta_0 \simeq 0.03$, while for 
$N=10^4$, we obtain $\eta_0 \simeq 4 \times 10^{-3}$. 

This result is not in contrast with previous work, such as \citep{Antonuccio1994}, since an Einstein-de Sitter (EdS, dust without curvature) model was assumed
there. In fact, for such a model, neglecting radiation, we have $H \propto a^{-3/2}$. In fact, \citep{Antonuccio1992} showed that it is possible to associate a relaxation timescale with dynamical friction in galaxy clusters, which then is of the order of the Hubble time, i.e., $1/\eta \propto H^{-1}$. $\eta$ can also be expressed in terms of the mass of the perturbation $M$. 
Recalling that $M=m_{\rm a}N$, it is straightforward to obtain
\begin{equation}\label{eqn:dyn_fric}
 \eta = 4.44\sqrt{\frac{3\Delta}{8\pi}}\frac{\log{[1.12(M/m_{\rm a})^{2/3}]}}{(M/m_{\rm a})}H = \eta_0\,H\,,
\end{equation}
where we assume $m_{\rm a}=10^9\,h^{-1}\,M_{\odot}$. It is evident that $\eta_0$ is smaller for massive objects than for low-mass perturbations. Eq.~\eqref{eqn:dyn_fric} comes from Eqs.~D3--D5 in \cite{DelPopolo09}, and namely from the theory of stochastic forces in a gravitational field, as shown by \cite{Kandrup1980}. The terms $m_a$ amd $n_a$ that determine
the value of $\eta$ are related to the theory of Gaussian stochastic fields, valid for generic structures like galaxies,
clusters, etc. In the text, we refer to the "protoclusters" just to estimate the values of these parameters. 
Again, in \citep{DelPopolo09}, the equations and values quoted are used to study clusters of galaxies.
Following \cite{Antonuccio1994}, one can show that the peaks of the local density field with central height $\nu\ge\nu_{\rm c}$, with $\nu_{\rm c}$ a critical threshold, contribute to dynamical friction. The number of these objects can be calculated under the condition that the peak radius $r_{\rm pk}(\nu\ge\nu_{\rm c})$ is negligible with respect to the average peak separation $n_{\rm a}^{-1/3}(\nu\ge\nu_{\rm c})$. The values required for the computation will necessarily depend on the particular cosmological model considered and on the filtering scale $R$. Here we do not pretend to derive exact values for the quantities above. Our aim is merely to estimate the order of magnitude of the relevant quantities, in particular the dynamical friction coefficient $\eta$ and to understand how this parameter affects the evolution of linear and nonlinear structures in $\Lambda$CDM and  models with more general DE.

\begin{figure*}[!ht]
 \centering
 \includegraphics[angle=0,width=0.43\hsize]{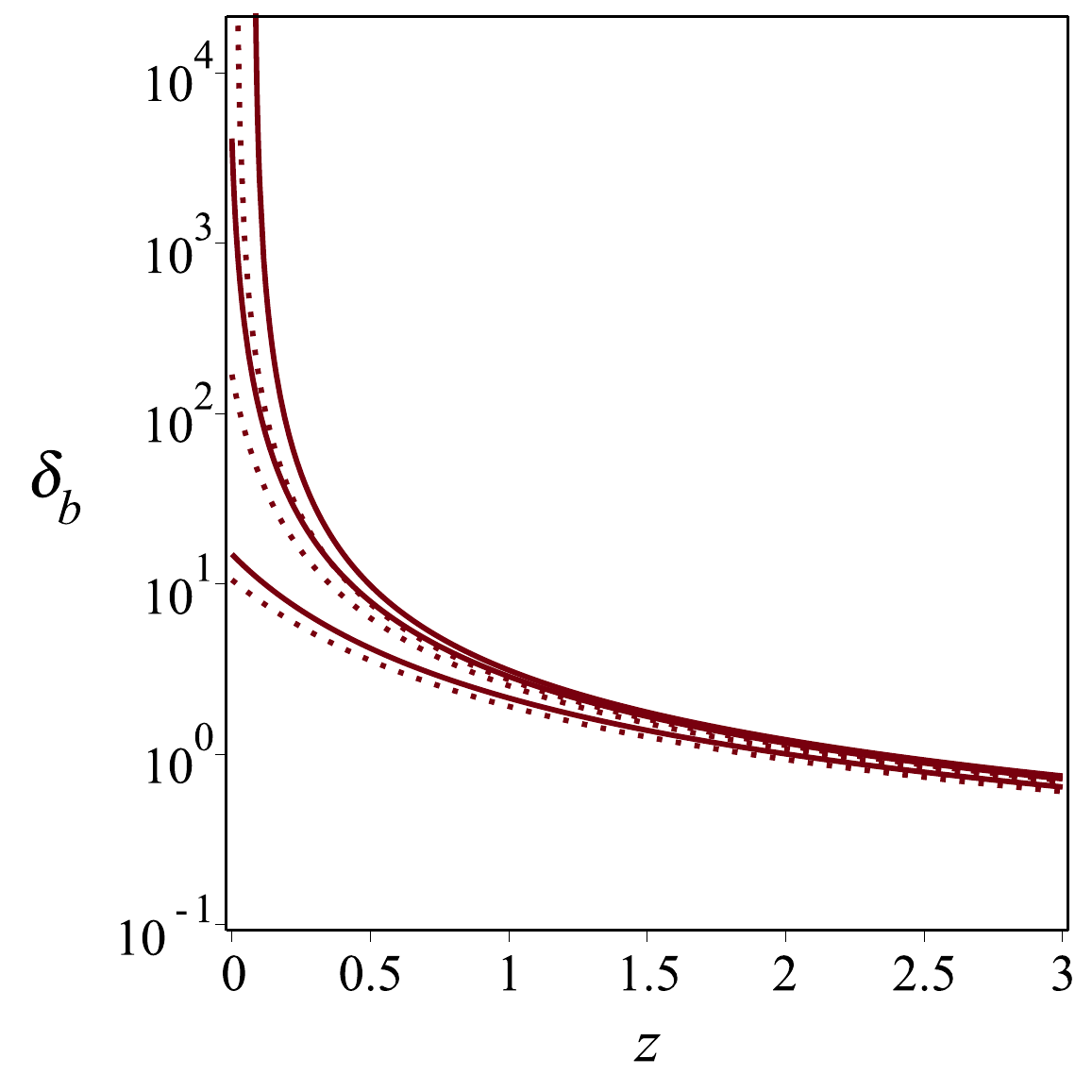}
 \includegraphics[angle=0,width=0.43\hsize]{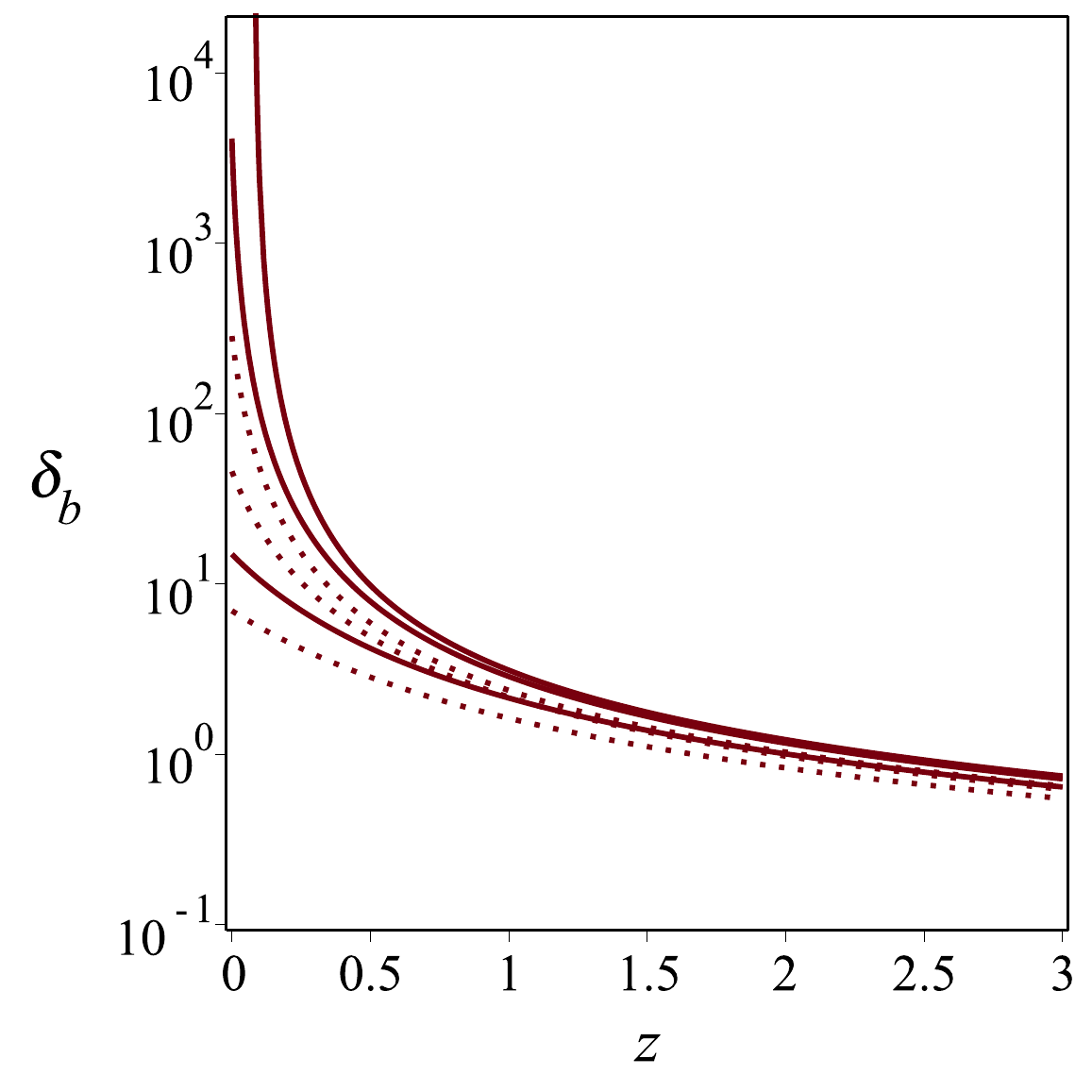} 
 \includegraphics[angle=0,width=0.43\hsize]{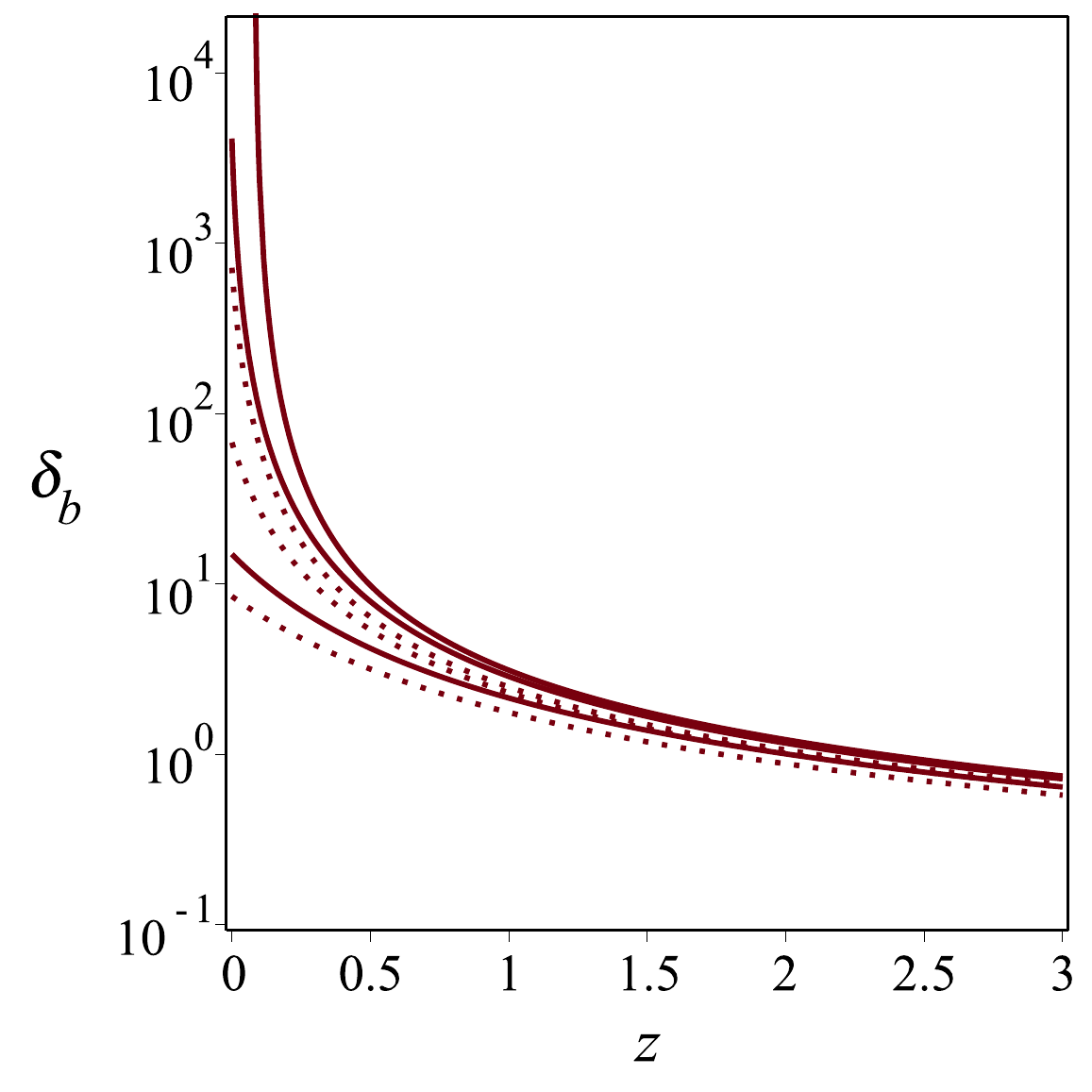} 
 \includegraphics[angle=0,width=0.43\hsize]{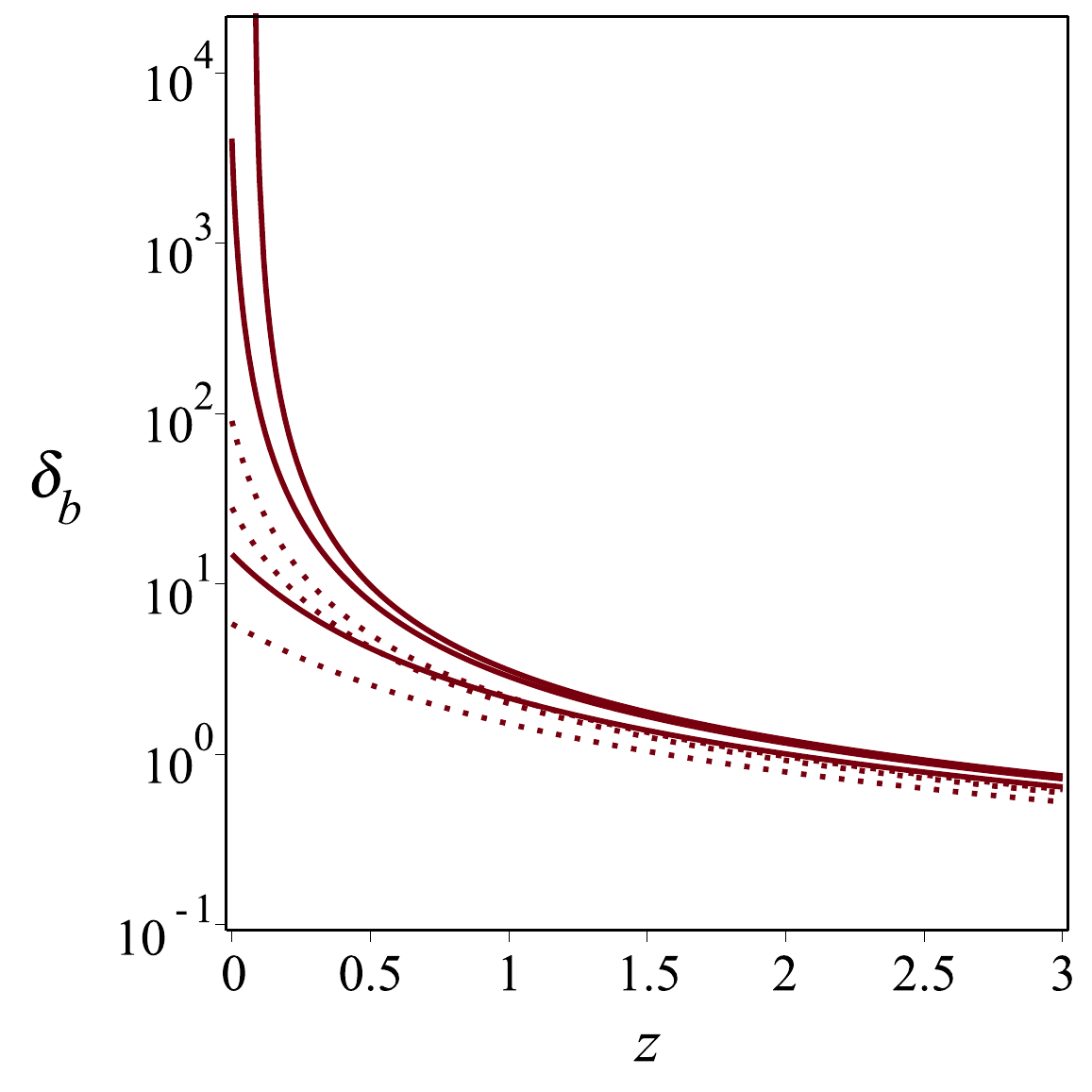} 
 \includegraphics[angle=0,width=0.43\hsize]{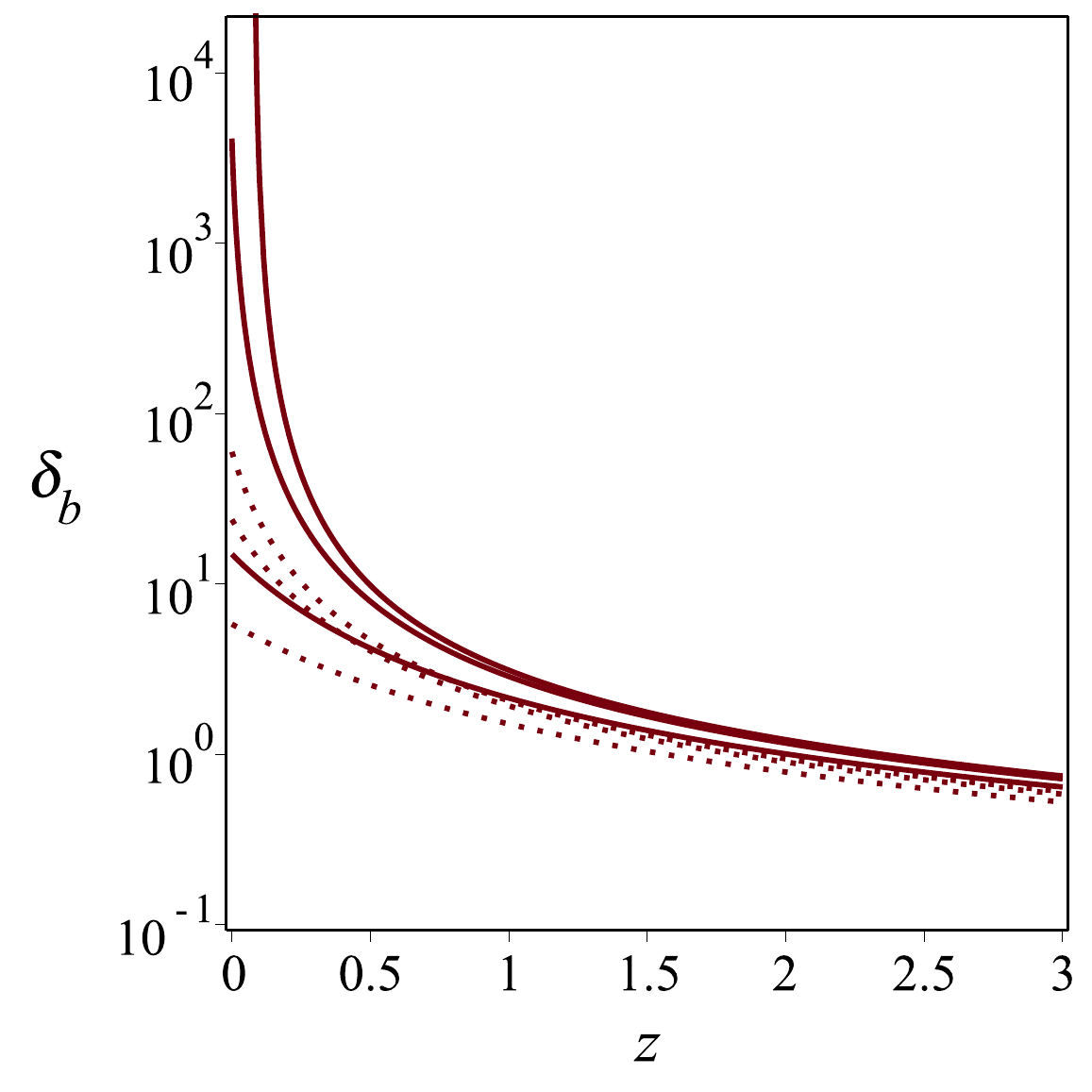} 
 \includegraphics[angle=0,width=0.43\hsize]{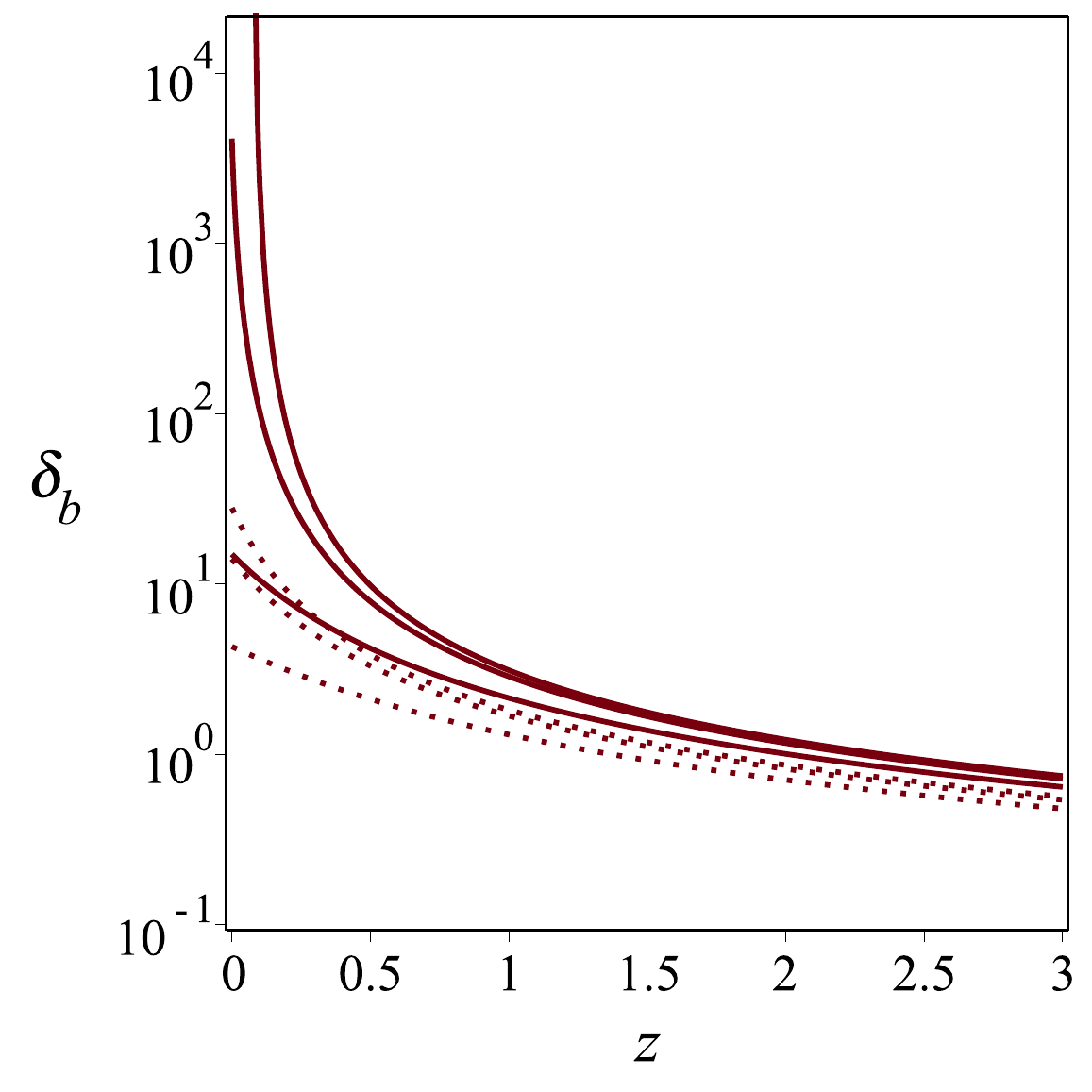}
 \caption{Growth of perturbations, $\delta_b$. The solid lines represents $\delta_b$
with $\alpha$ increasing from 0 (bottom solid lines), to 0.5 (central solid lines), and 1 (top solid lines), for $\beta=0$. The dotted lines represent $\delta_b$, with $\beta = 0.01$ (top two panels), $\beta=0.02$ (central two panels), and $\beta=0.04$ (bottom two panels). The value of $\eta_0$ is equal to $4 \times 10^{-3}$ in all left column panels, and 0.03 in all right column panels.}
\label{fig:deltas_vs_z1}
\end{figure*}

\begin{figure*}[!ht]
 \centering
 \includegraphics[angle=0,width=0.43\hsize]{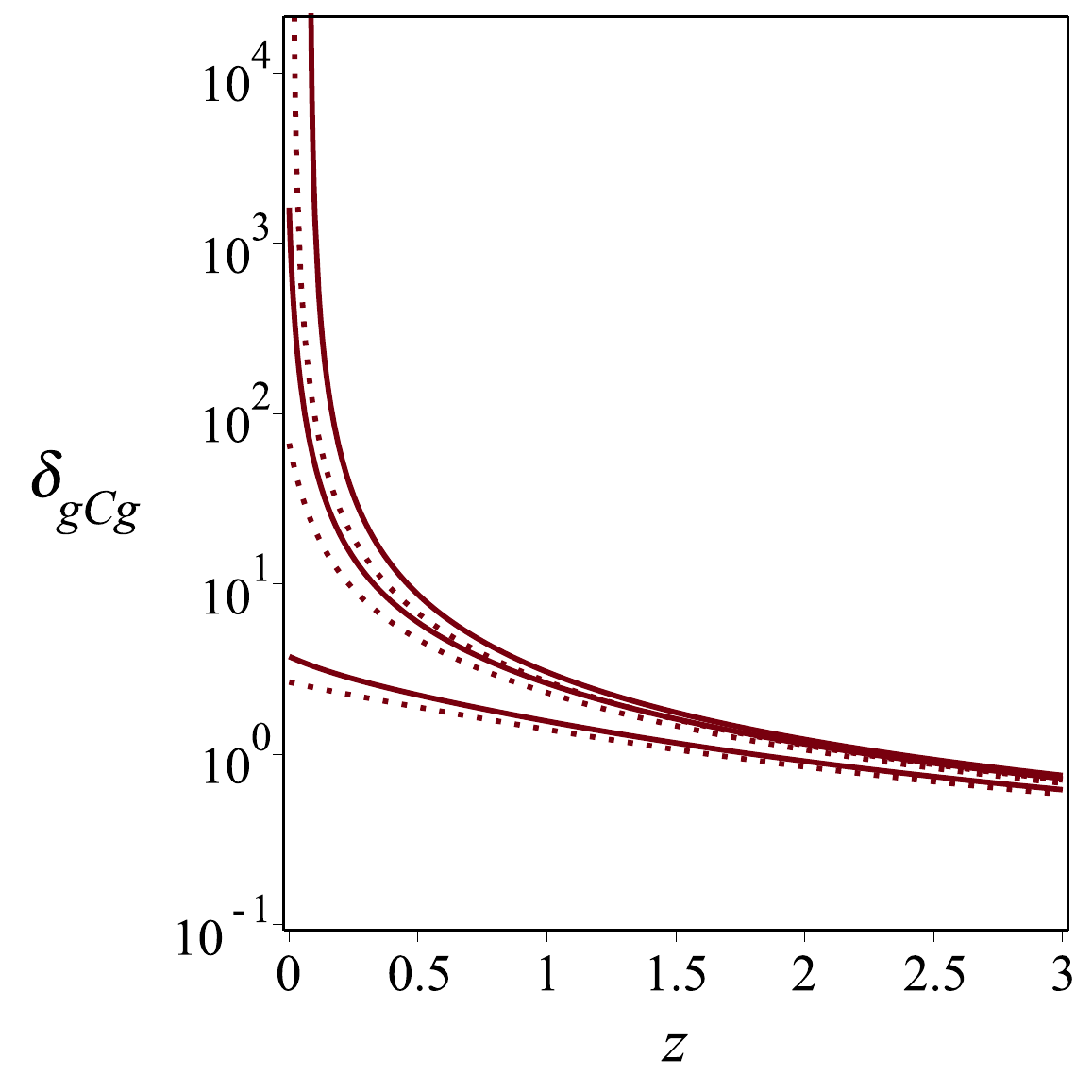}
 \includegraphics[angle=0,width=0.43\hsize]{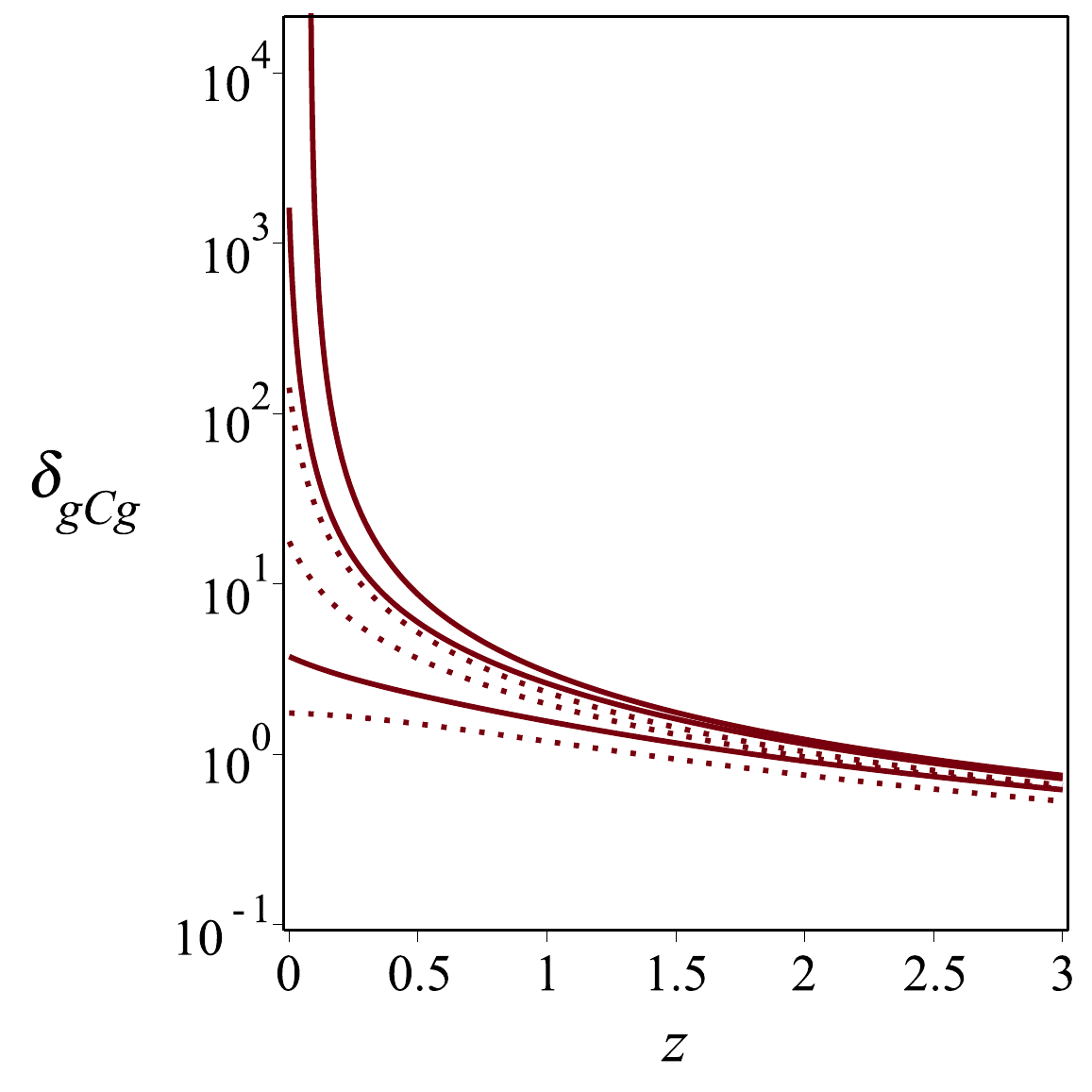} 
 \includegraphics[angle=0,width=0.43\hsize]{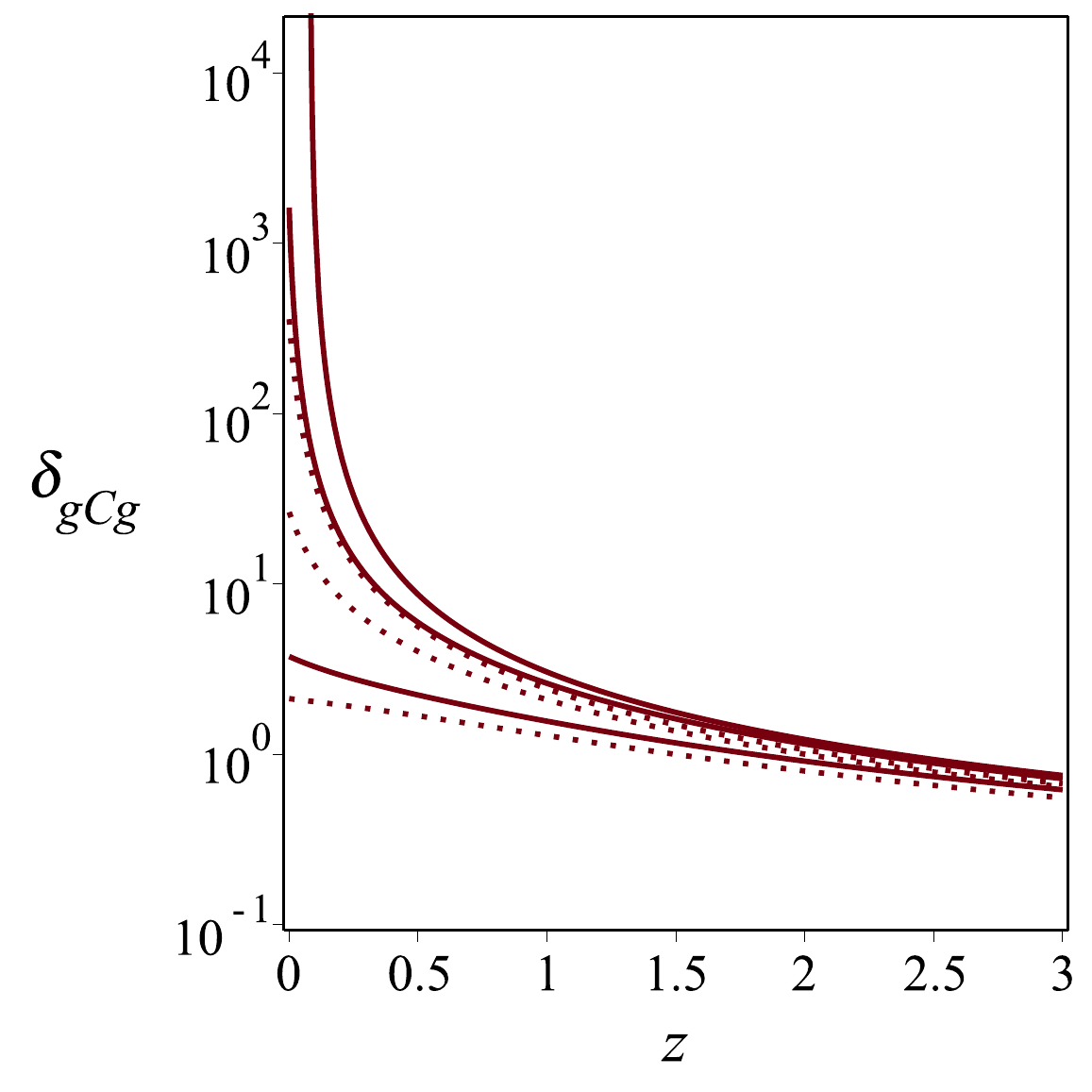} 
 \includegraphics[angle=0,width=0.43\hsize]{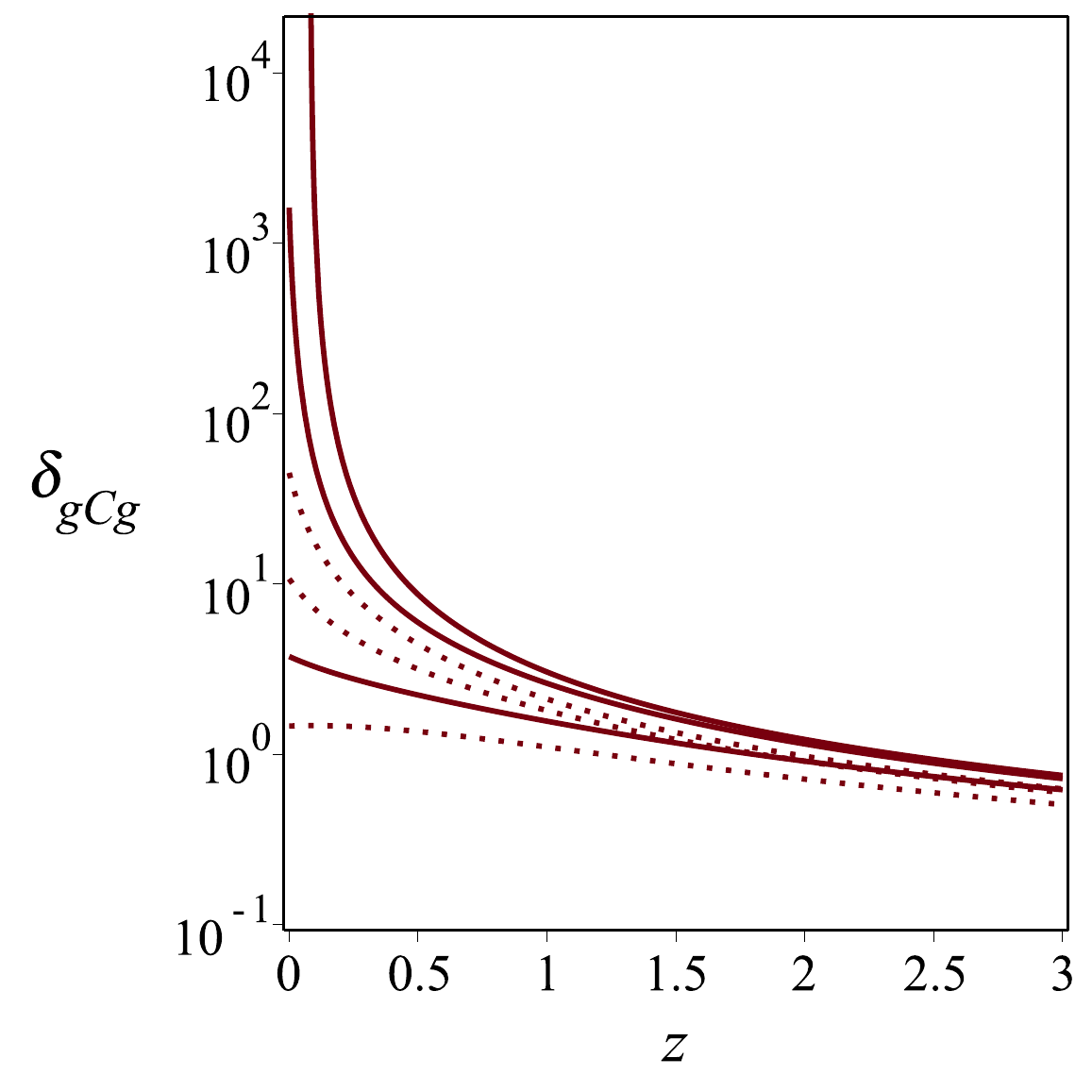} 
 \includegraphics[angle=0,width=0.43\hsize]{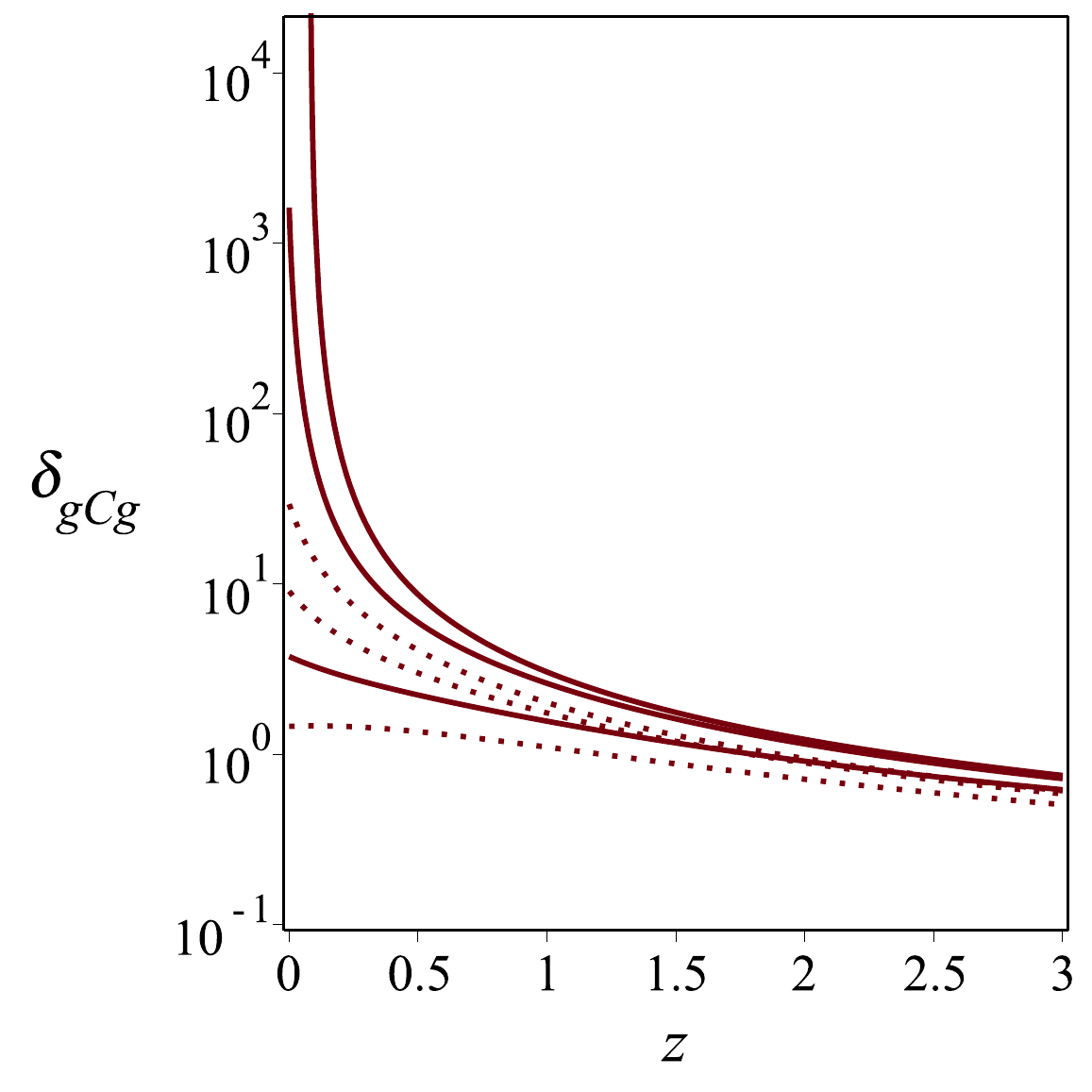} 
 \includegraphics[angle=0,width=0.43\hsize]{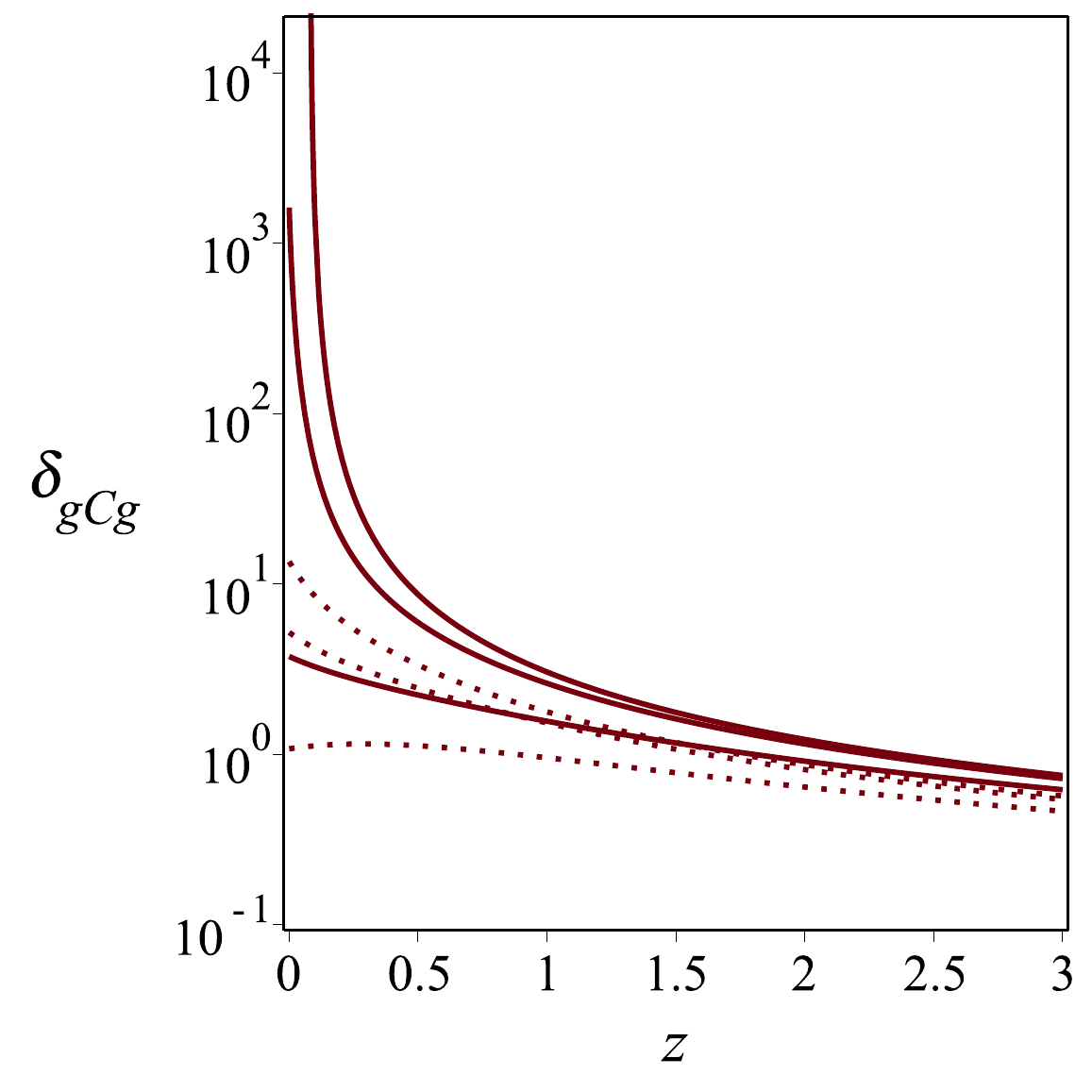}
 \caption{Growth of perturbations, $\delta_{GCG}$. Lines description is as in Fig.~\ref{fig:deltas_vs_z1}.}
\label{fig:deltas_vs_z2}
\end{figure*}

\begin{figure*}[!ht]
 \centering
 \includegraphics[angle=0,width=0.43\hsize]{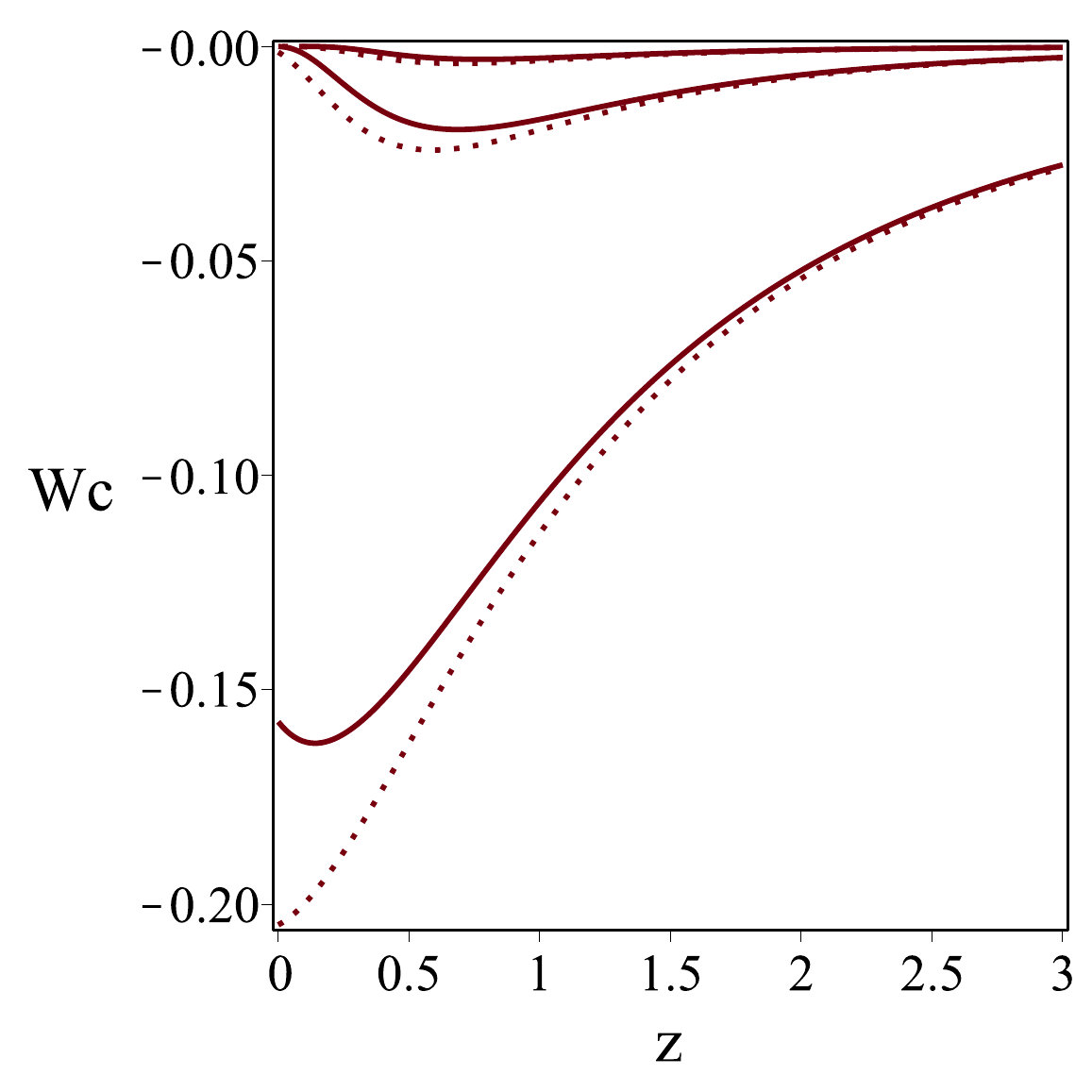}
 \includegraphics[angle=0,width=0.43\hsize]{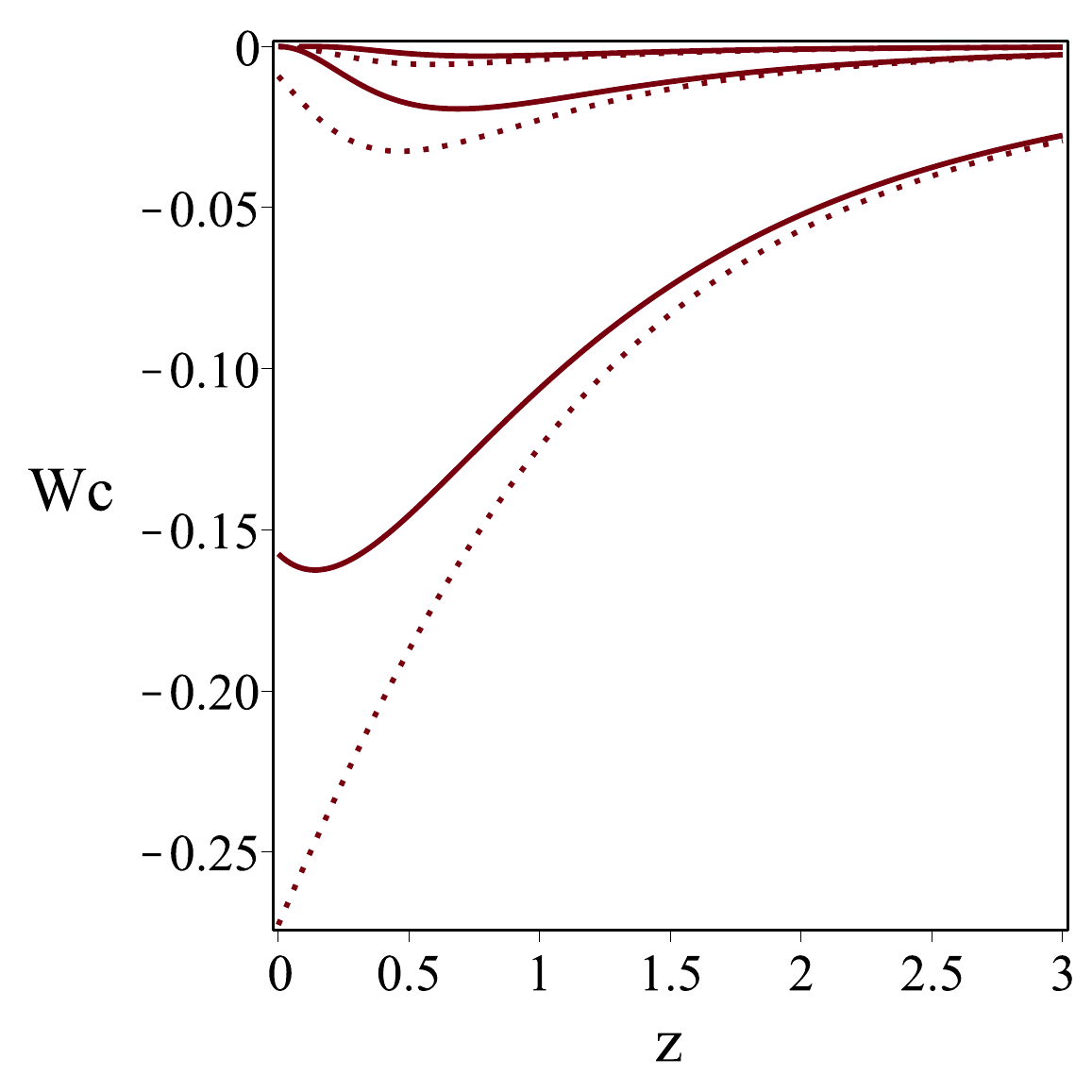} 
 \includegraphics[angle=0,width=0.43\hsize]{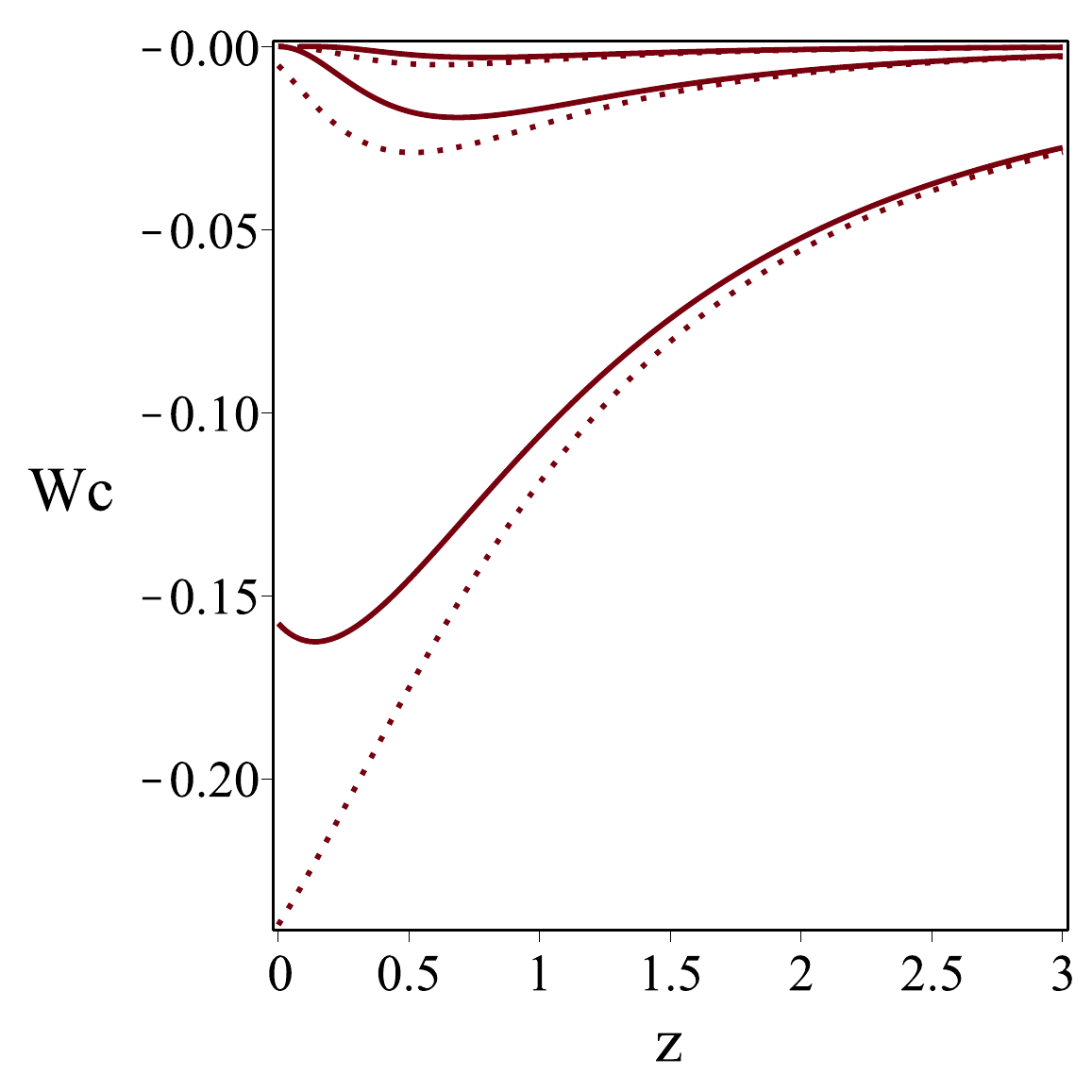} 
 \includegraphics[angle=0,width=0.43\hsize]{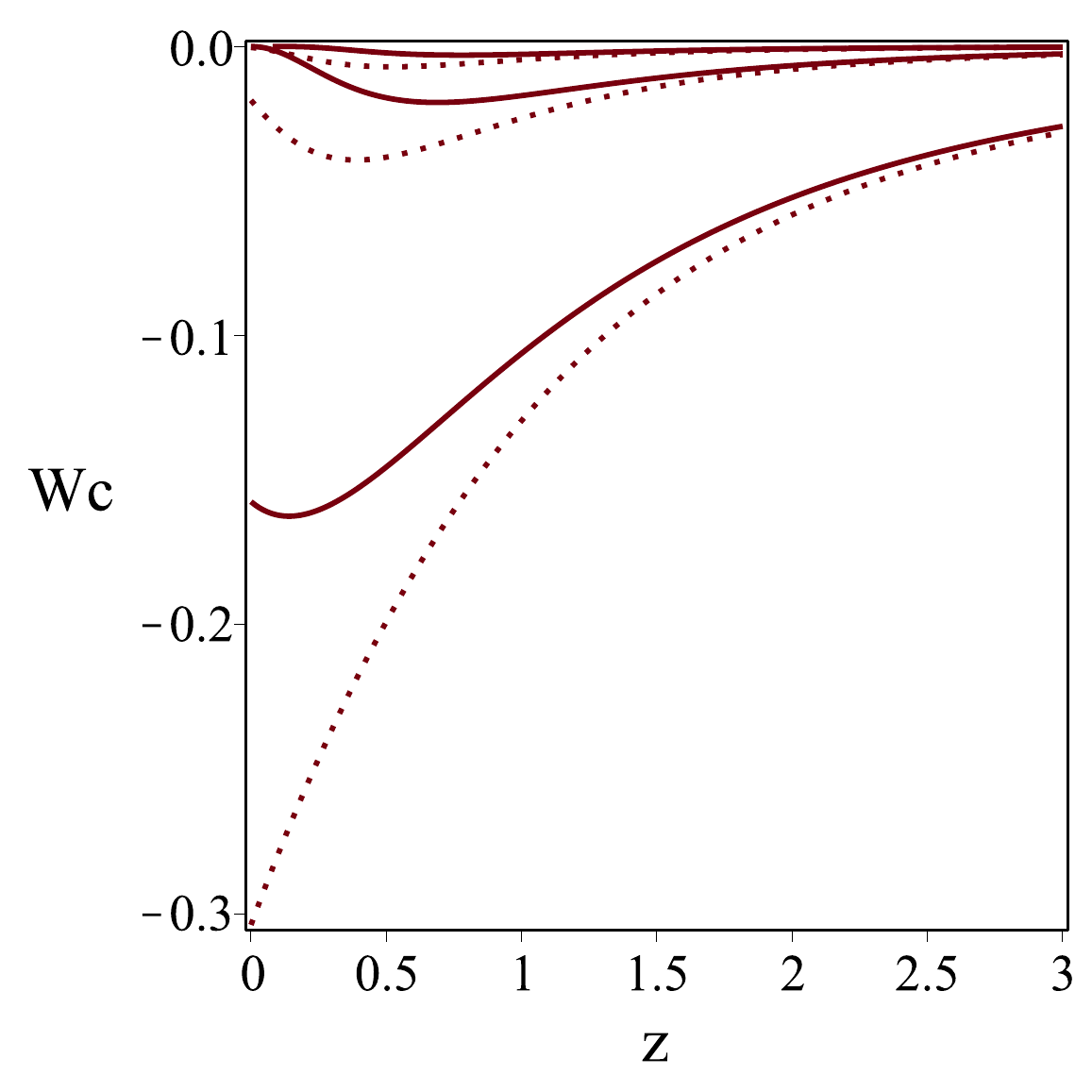} 
 \includegraphics[angle=0,width=0.43\hsize]{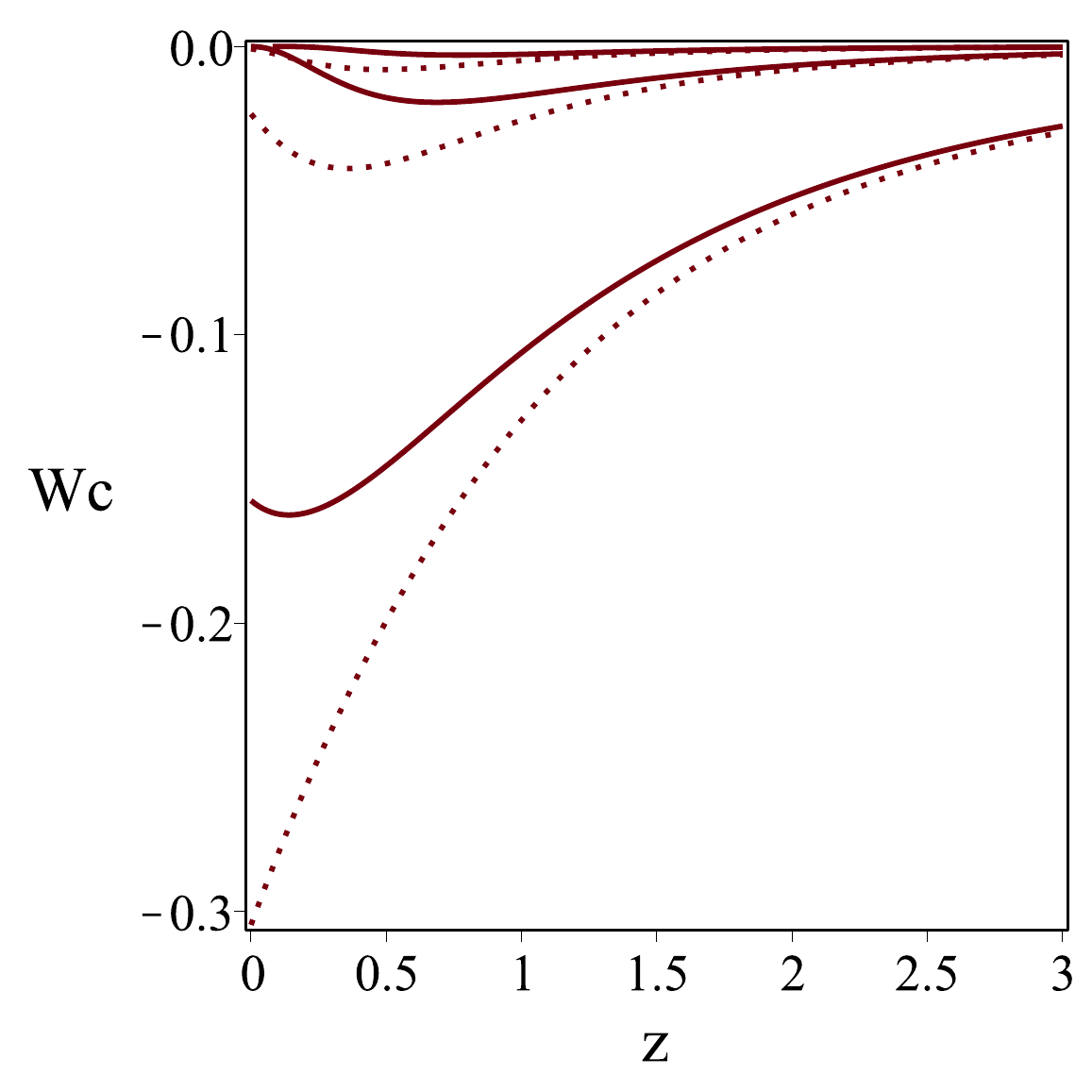} 
 \includegraphics[angle=0,width=0.43\hsize]{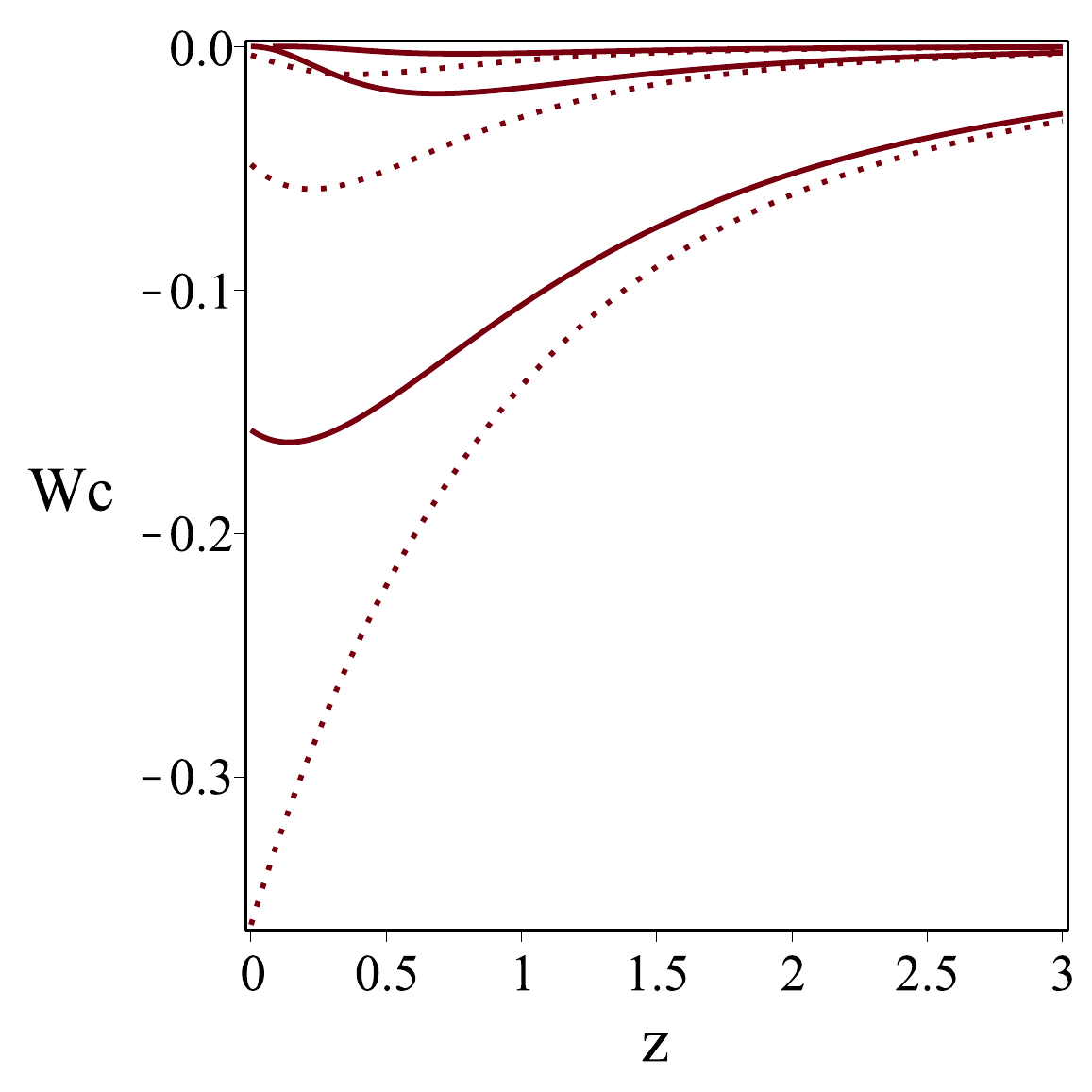}
 \caption{Evolution of $w_c$ with $z$ for GCG universes. Lines description is as in previous Figs.~\ref{fig:deltas_vs_z1}--\ref{fig:deltas_vs_z2}.}
\label{fig:deltas_vs_z3}
\end{figure*}

\begin{figure*}[!ht]
 \centering
 \includegraphics[angle=0,width=0.43\hsize]{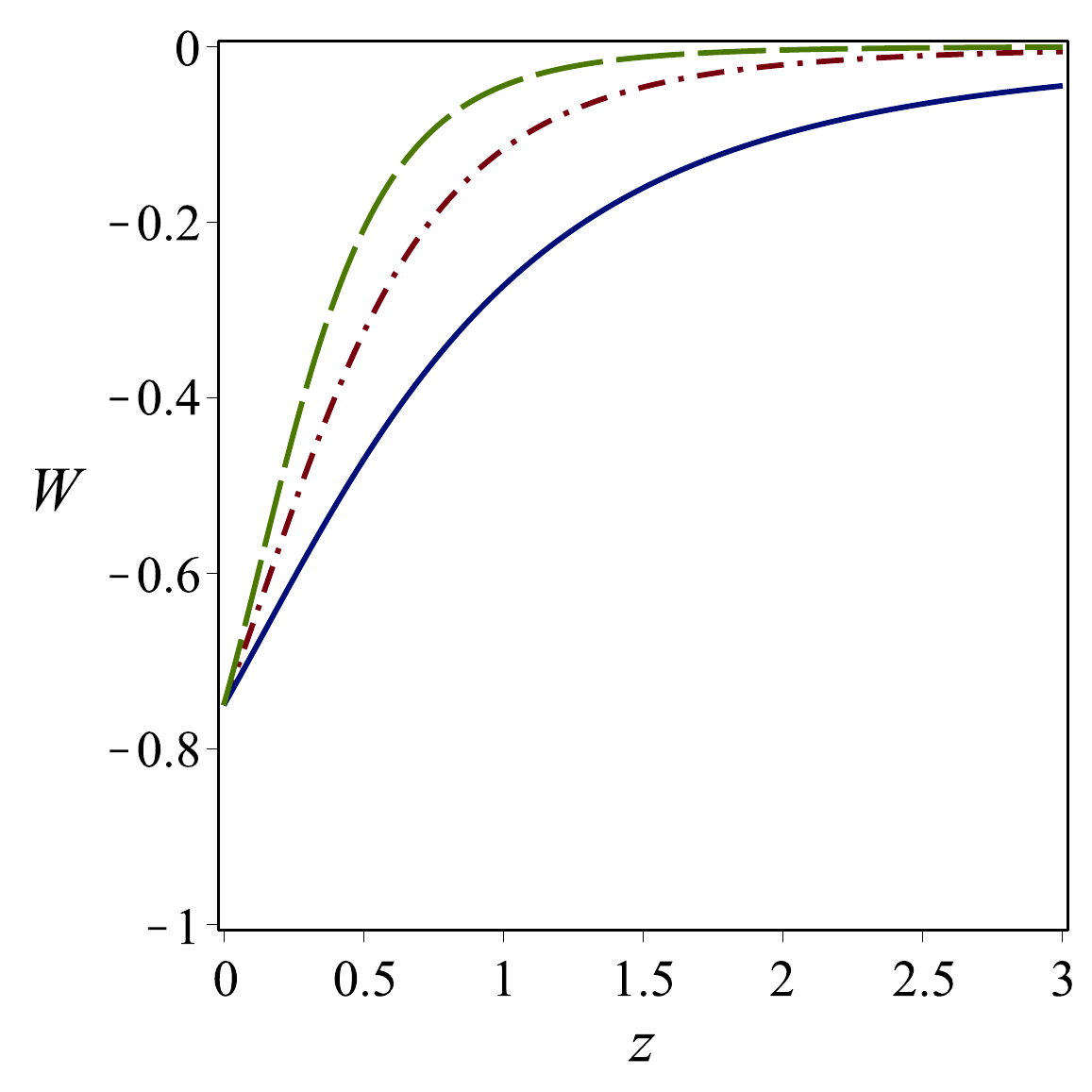}
 \caption{Evolution of $w$ with $z$ for GCG universes for $\alpha = 1$ (dashed line), 0.5 (dot-dashed line), 0 (solid line).}
\label{fig:deltas_vs_z4}
\end{figure*}

\begin{figure*}[!ht]
 \centering
 \includegraphics[angle=0,width=0.43\hsize]{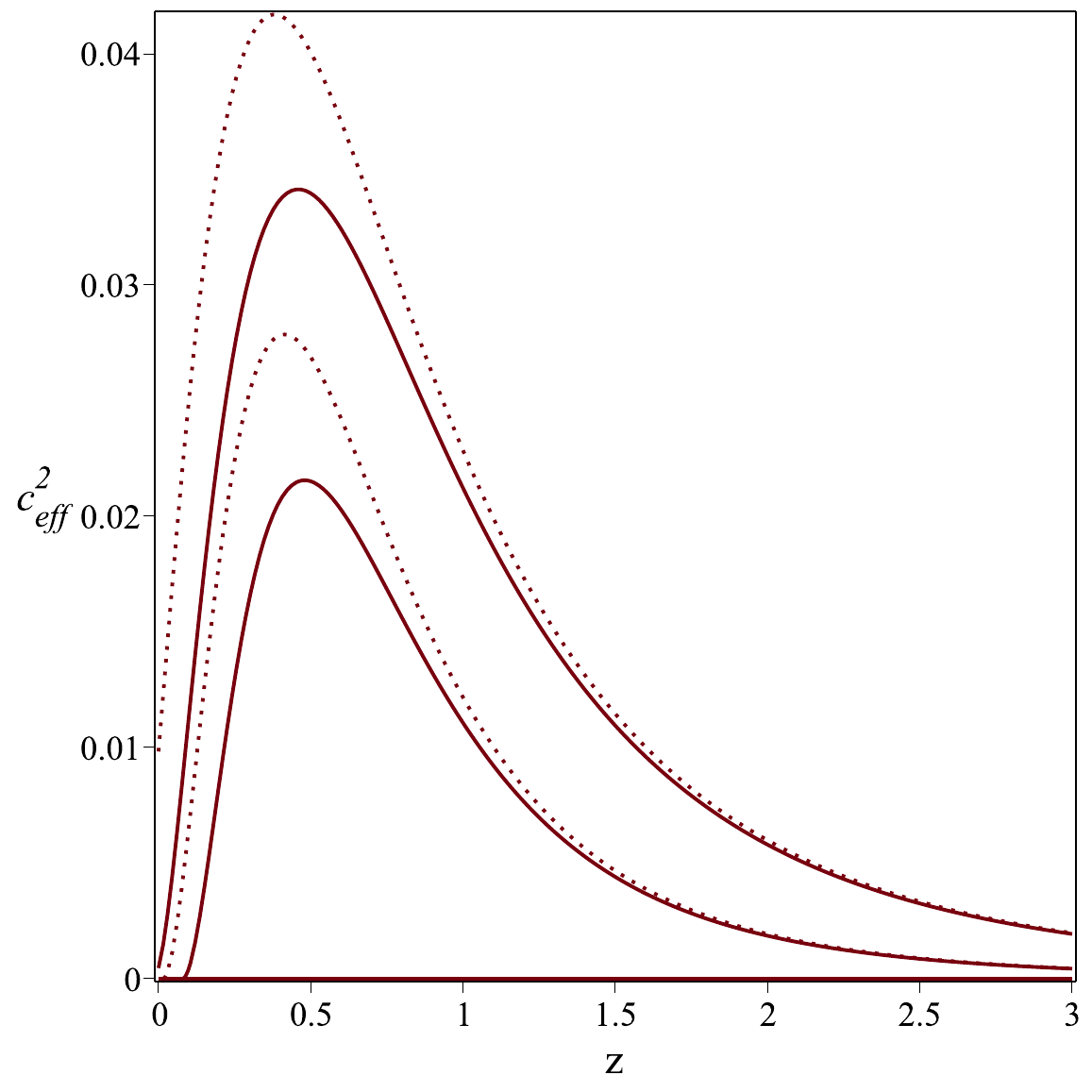}
 \includegraphics[angle=0,width=0.43\hsize]{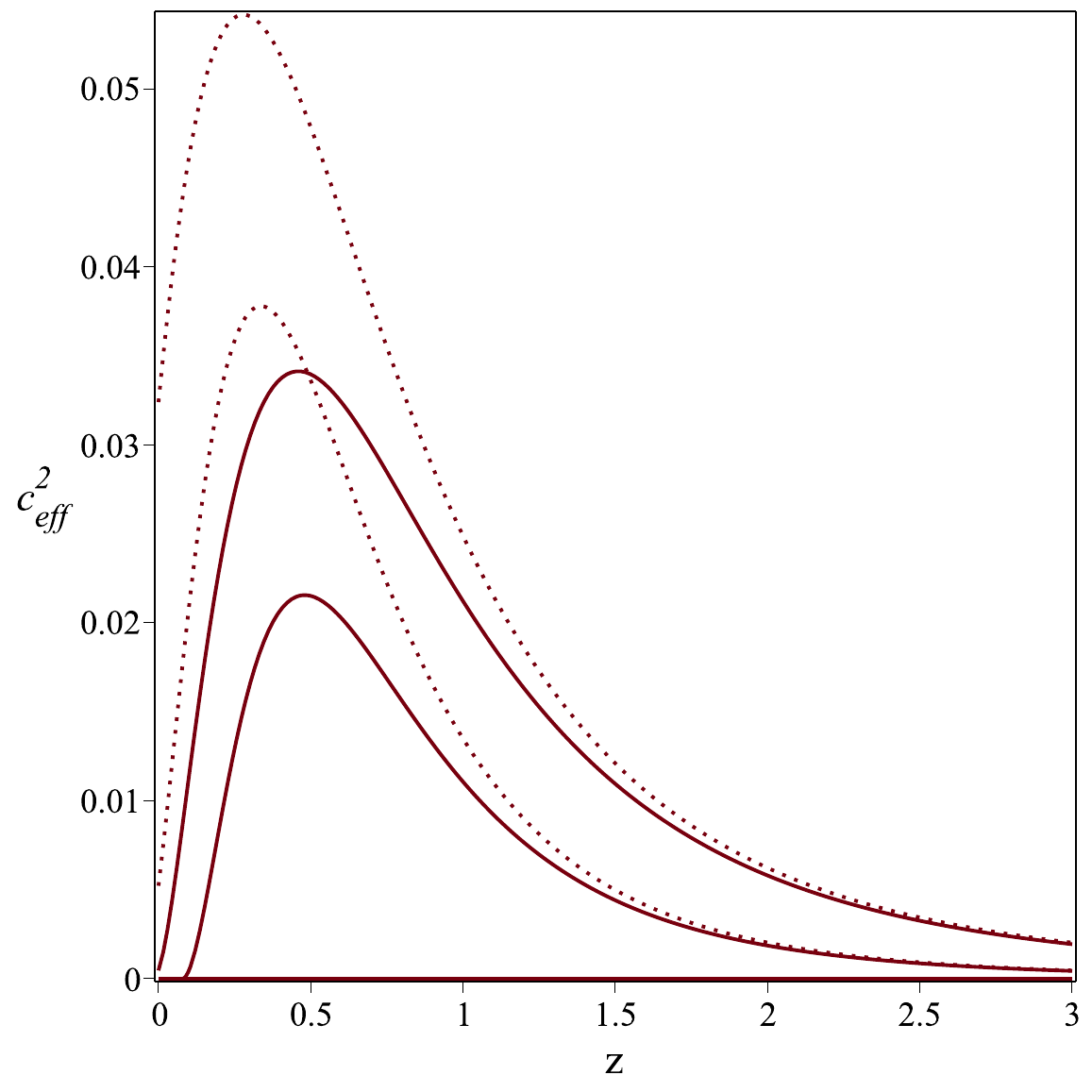} 
 \includegraphics[angle=0,width=0.43\hsize]{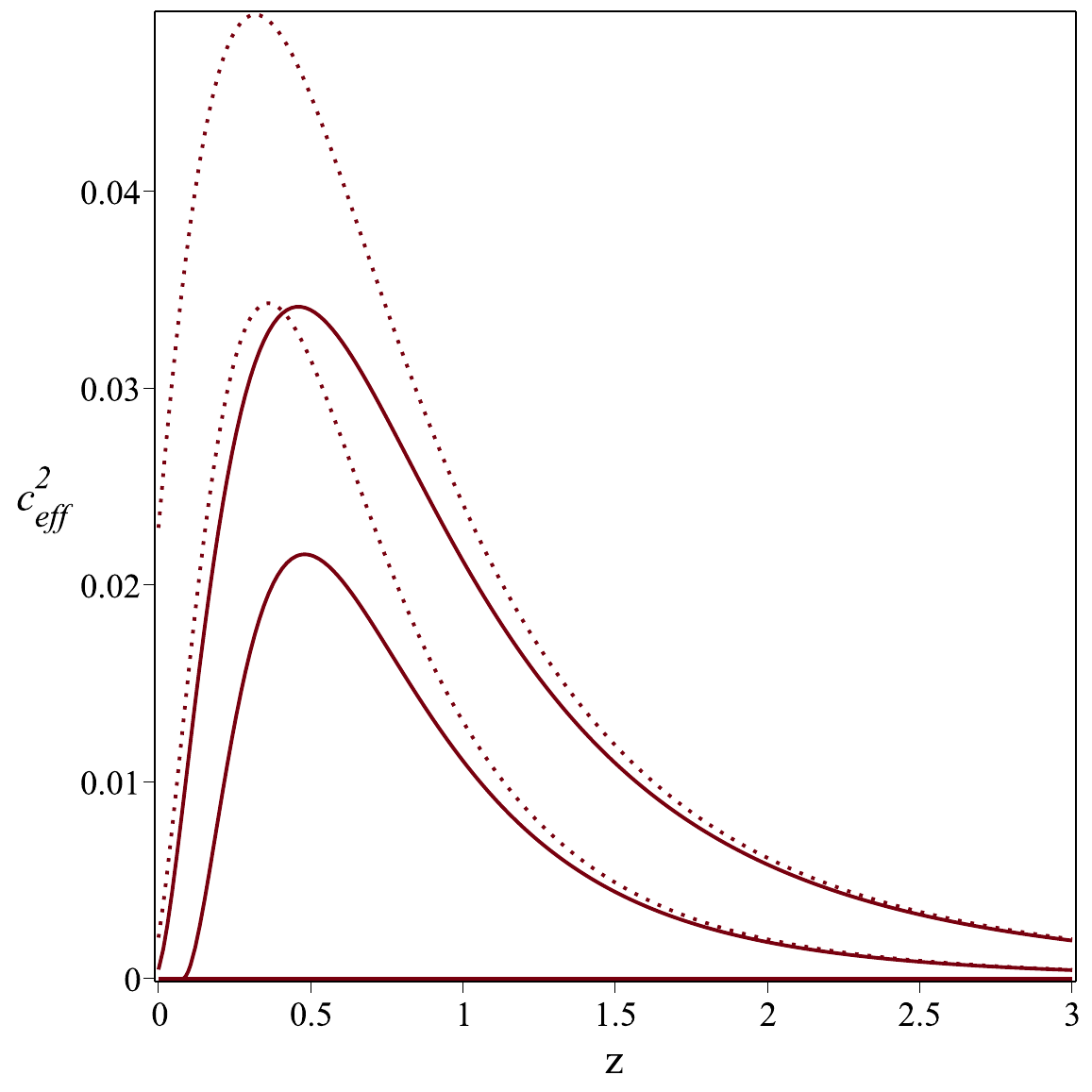} 
 \includegraphics[angle=0,width=0.43\hsize]{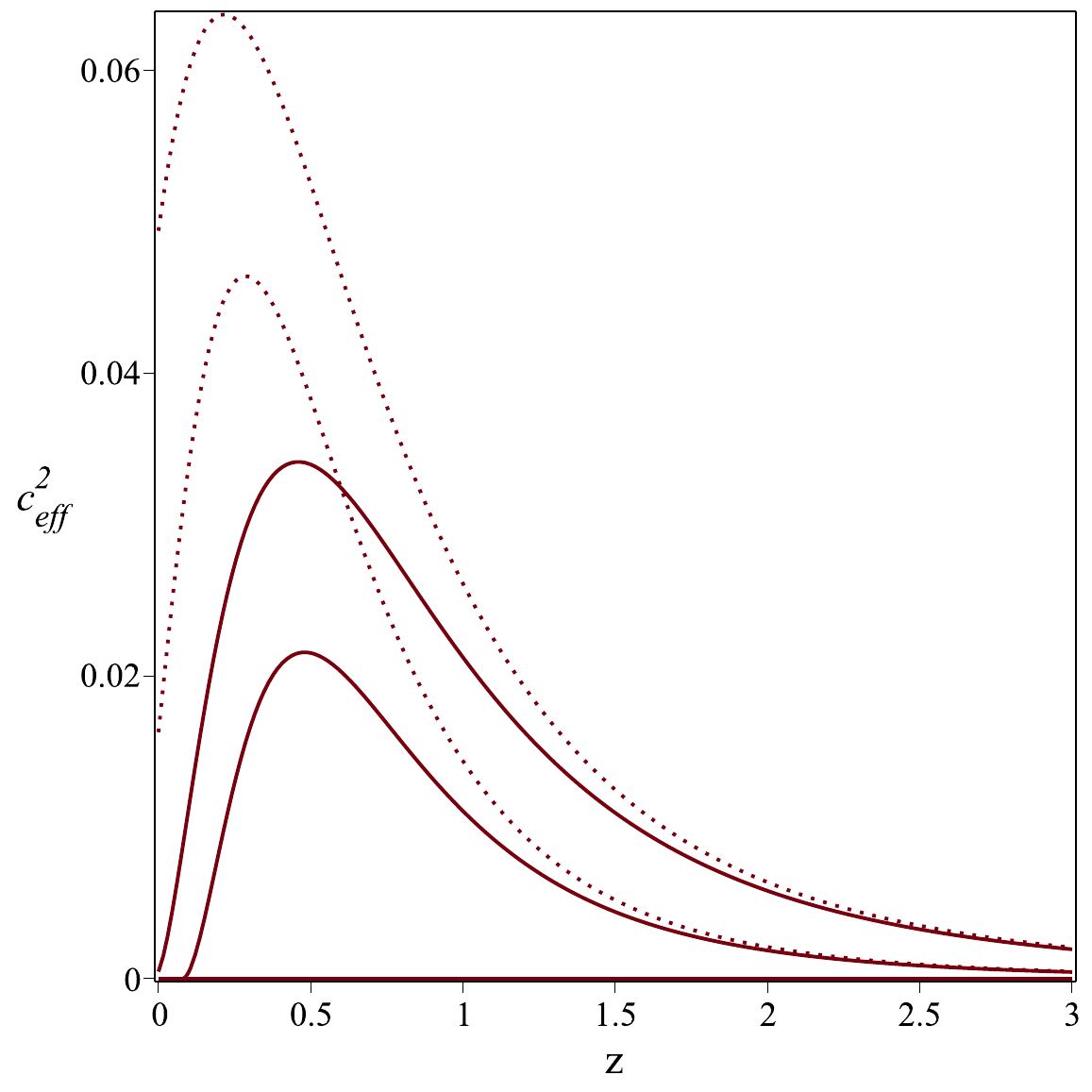} 
 \includegraphics[angle=0,width=0.43\hsize]{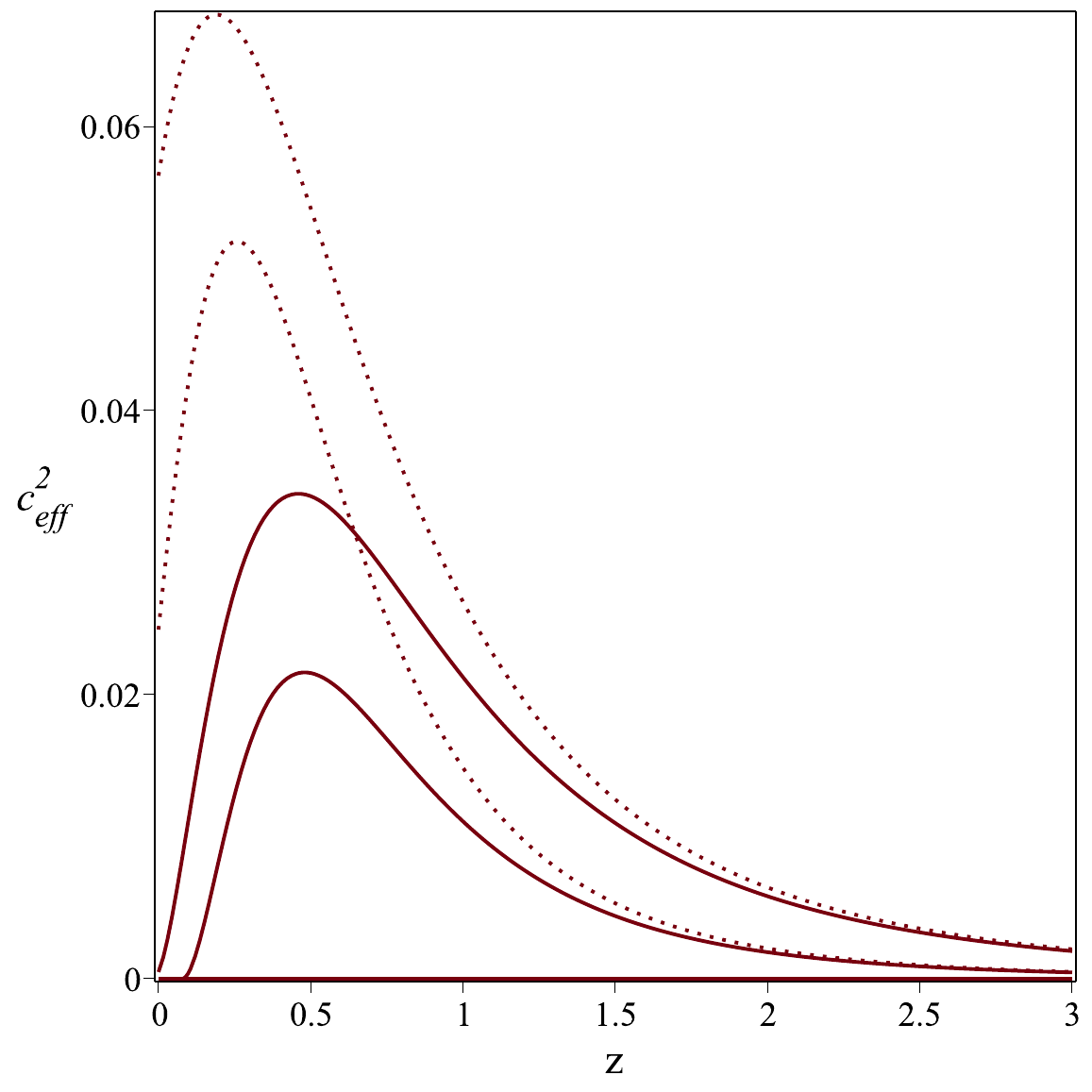} 
 \includegraphics[angle=0,width=0.43\hsize]{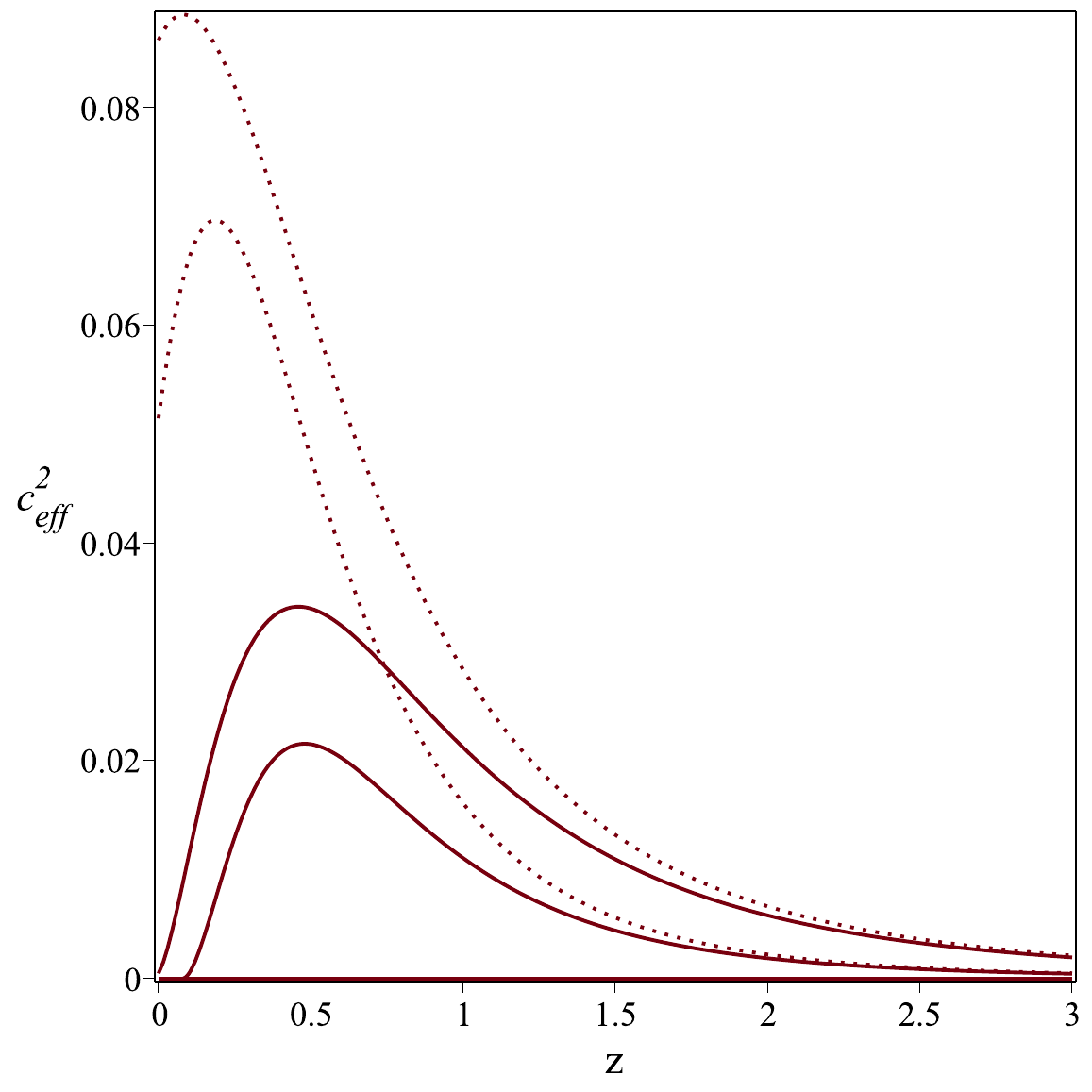}
 \caption{Evolution of $c^2_{\rm eff}$ with $z$ for GCG universes. Lines description is as in previous Figs.~\ref{fig:deltas_vs_z1}--\ref{fig:deltas_vs_z3}.}
\label{fig:deltas_vs_z5}
\end{figure*}

\begin{figure*}[!ht]
 \centering
 \includegraphics[angle=0,width=0.43\hsize]{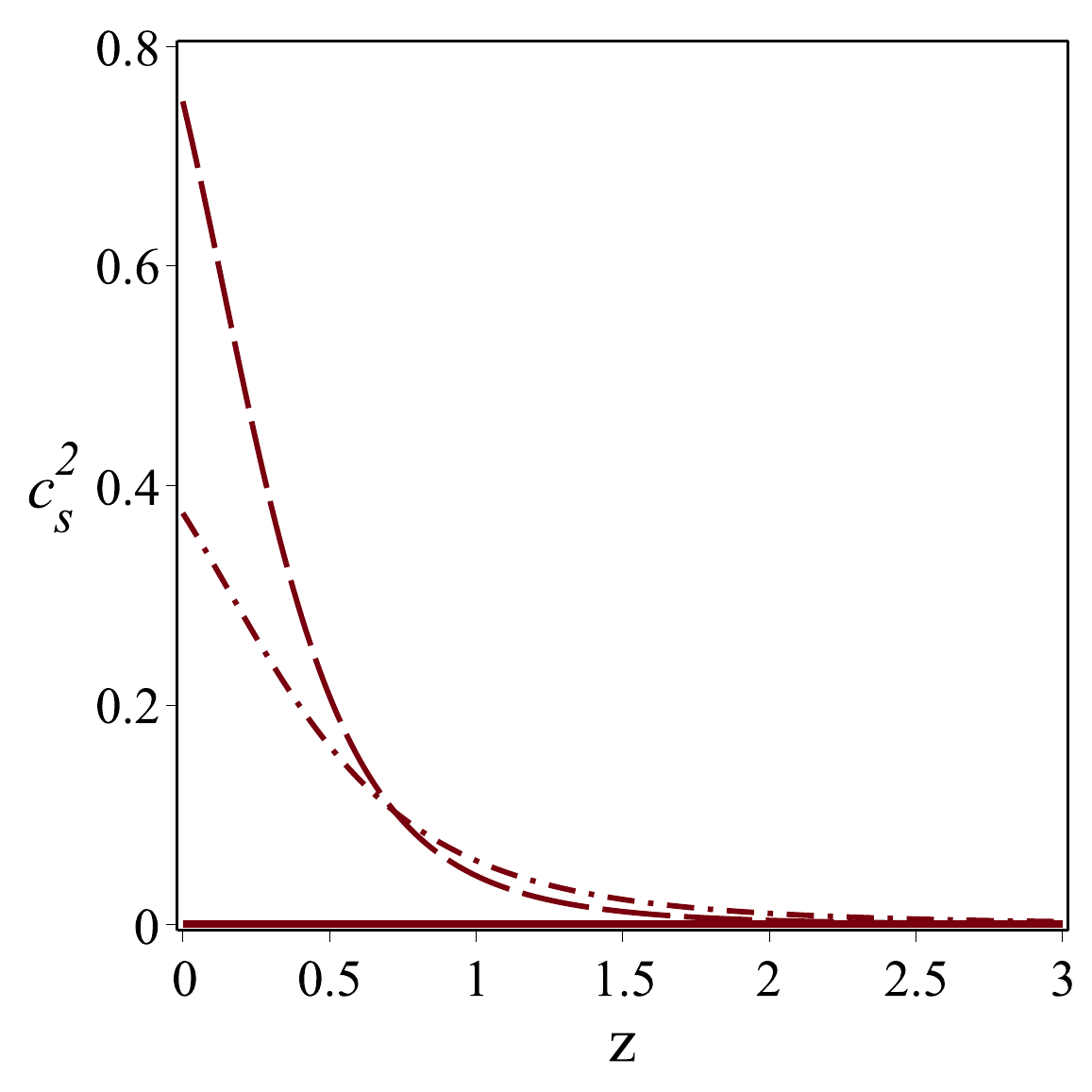}
 \caption{Evolution of $c^2_{\rm s}$ with $z$ for GCG universes for $\alpha = 1$ (dashed line), 0.5 (dot-dashed line), 0 (solid line).}
\label{fig:deltas_vs_z6}
\end{figure*}

\begin{figure*}[!ht]
 \centering
 \includegraphics[angle=0,width=0.43\hsize]{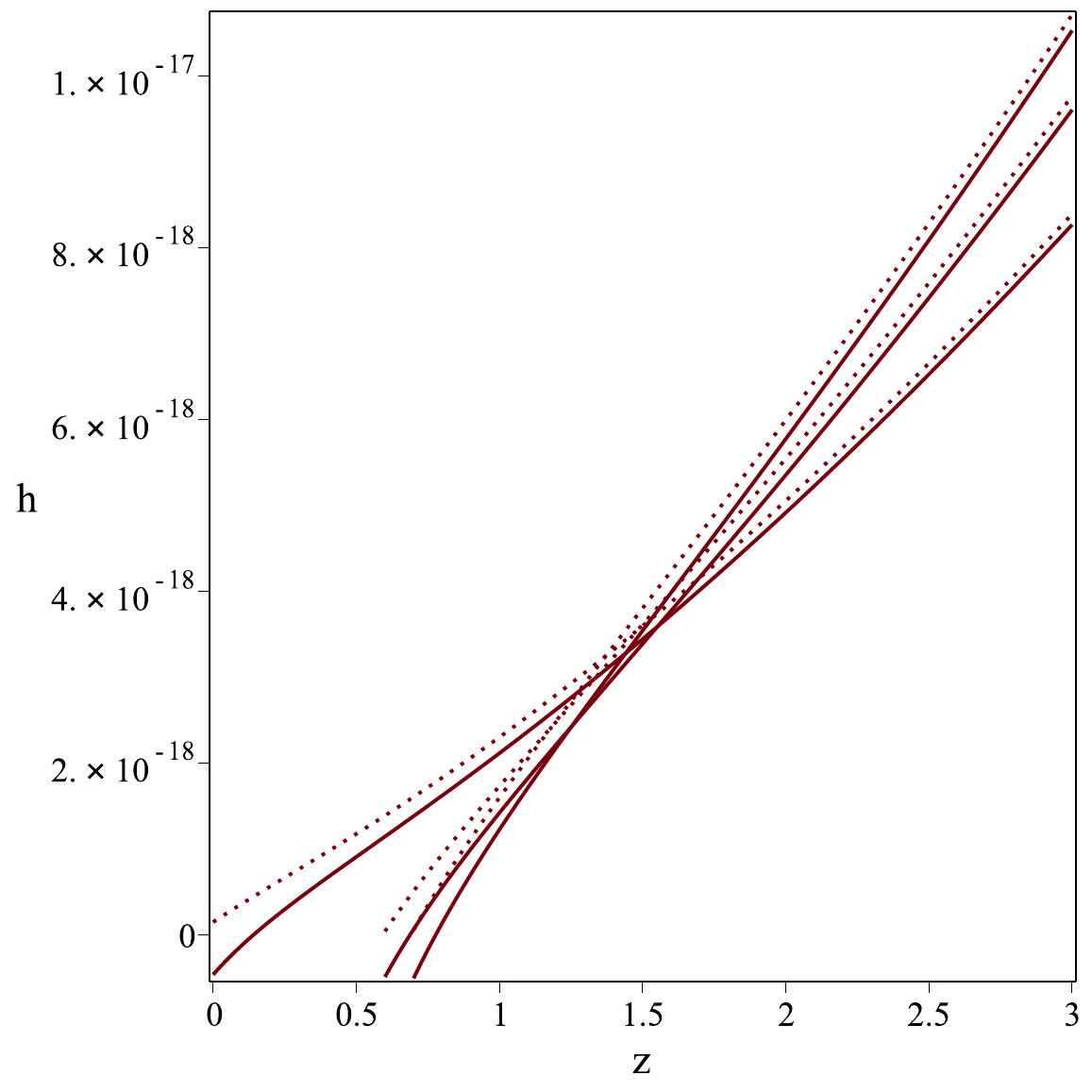}
 \includegraphics[angle=0,width=0.43\hsize]{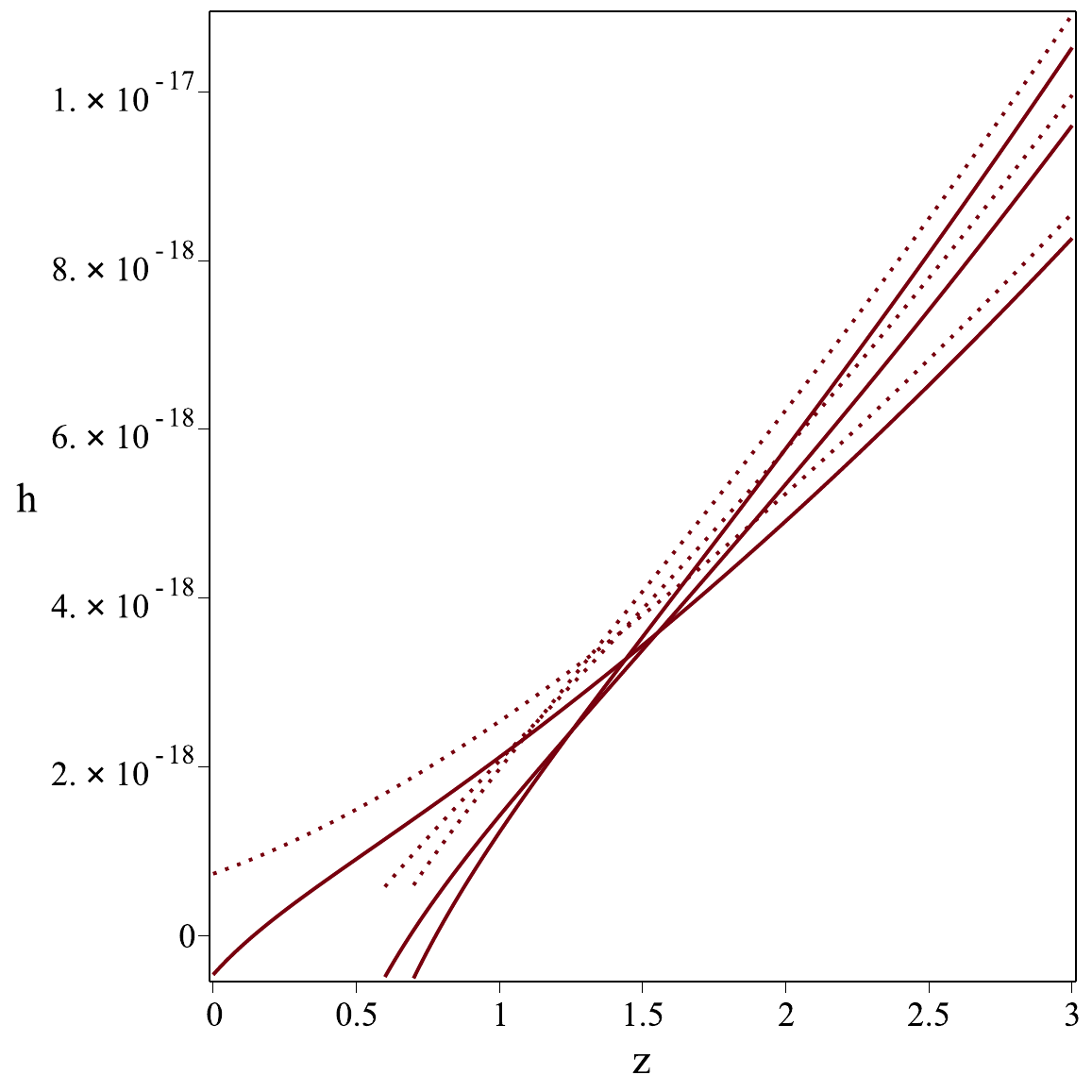} 
 \includegraphics[angle=0,width=0.43\hsize]{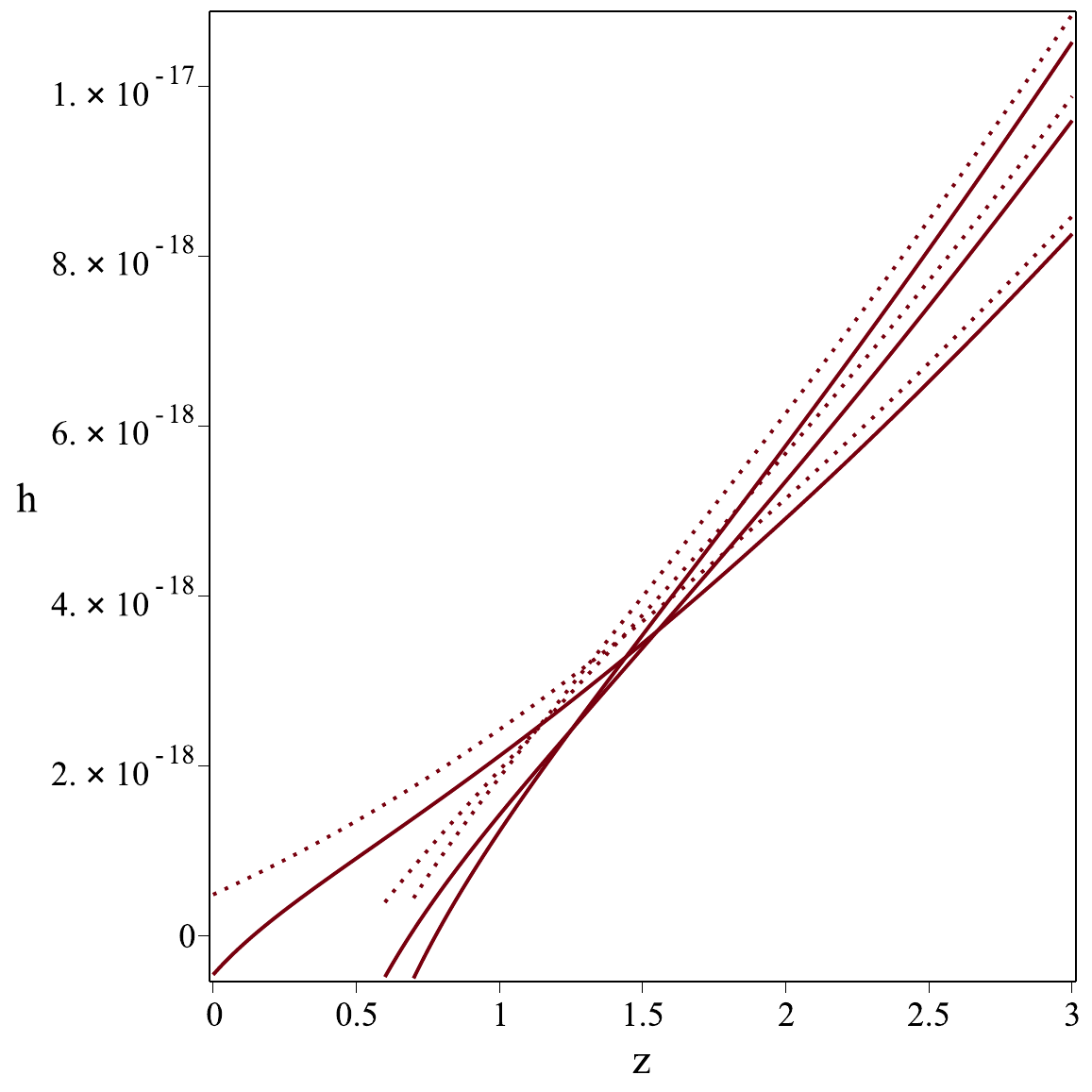} 
 \includegraphics[angle=0,width=0.43\hsize]{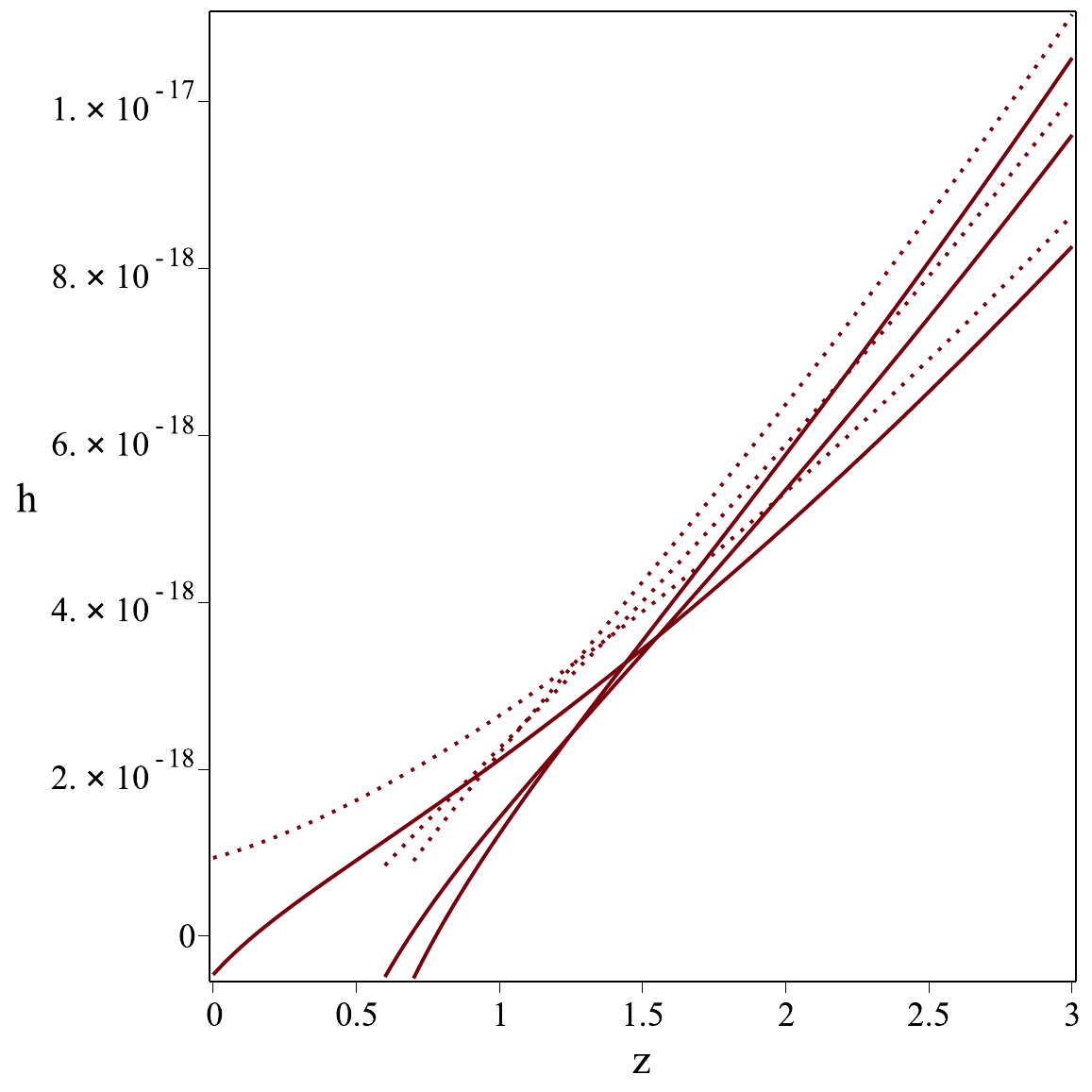} 
 \includegraphics[angle=0,width=0.43\hsize]{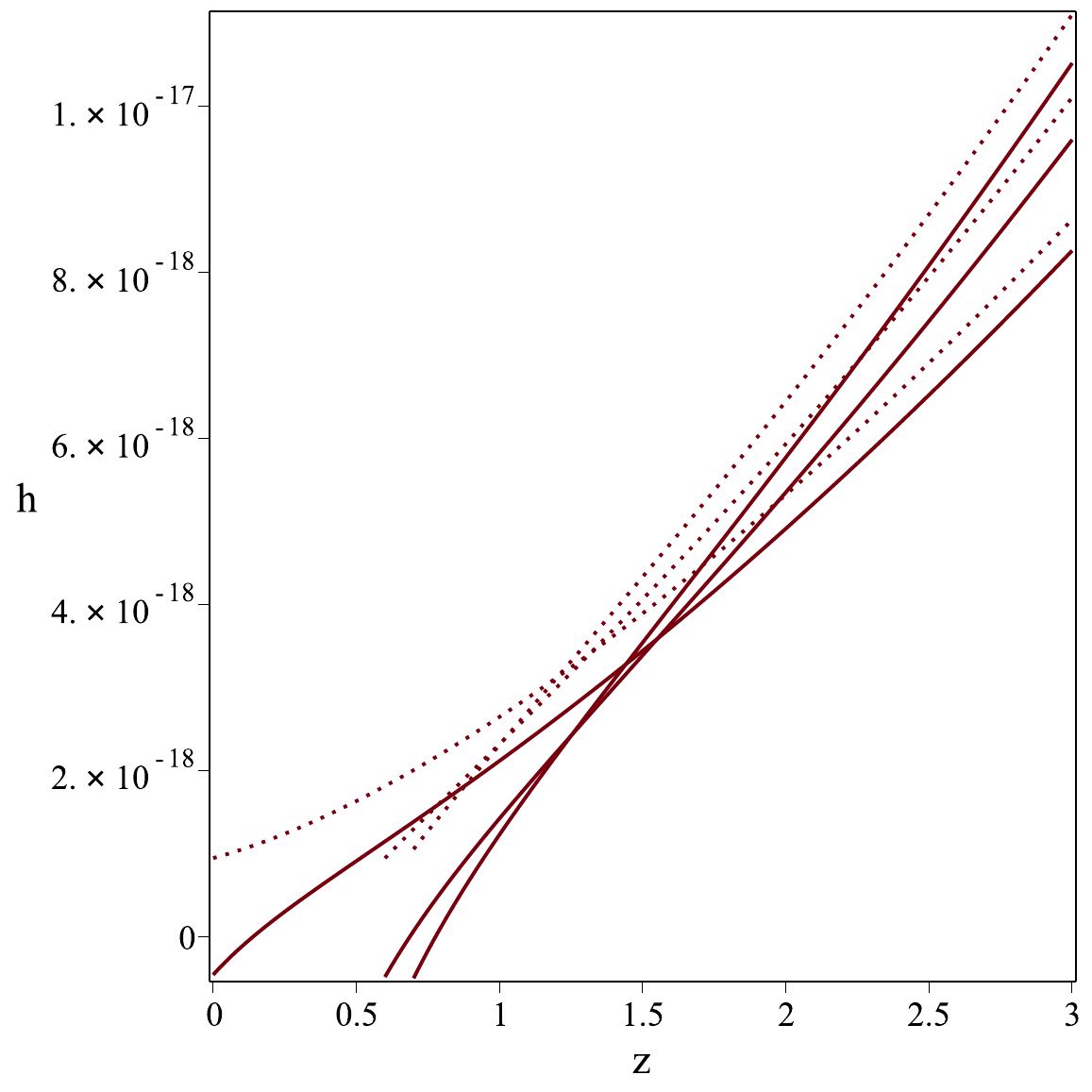} 
 \includegraphics[angle=0,width=0.43\hsize]{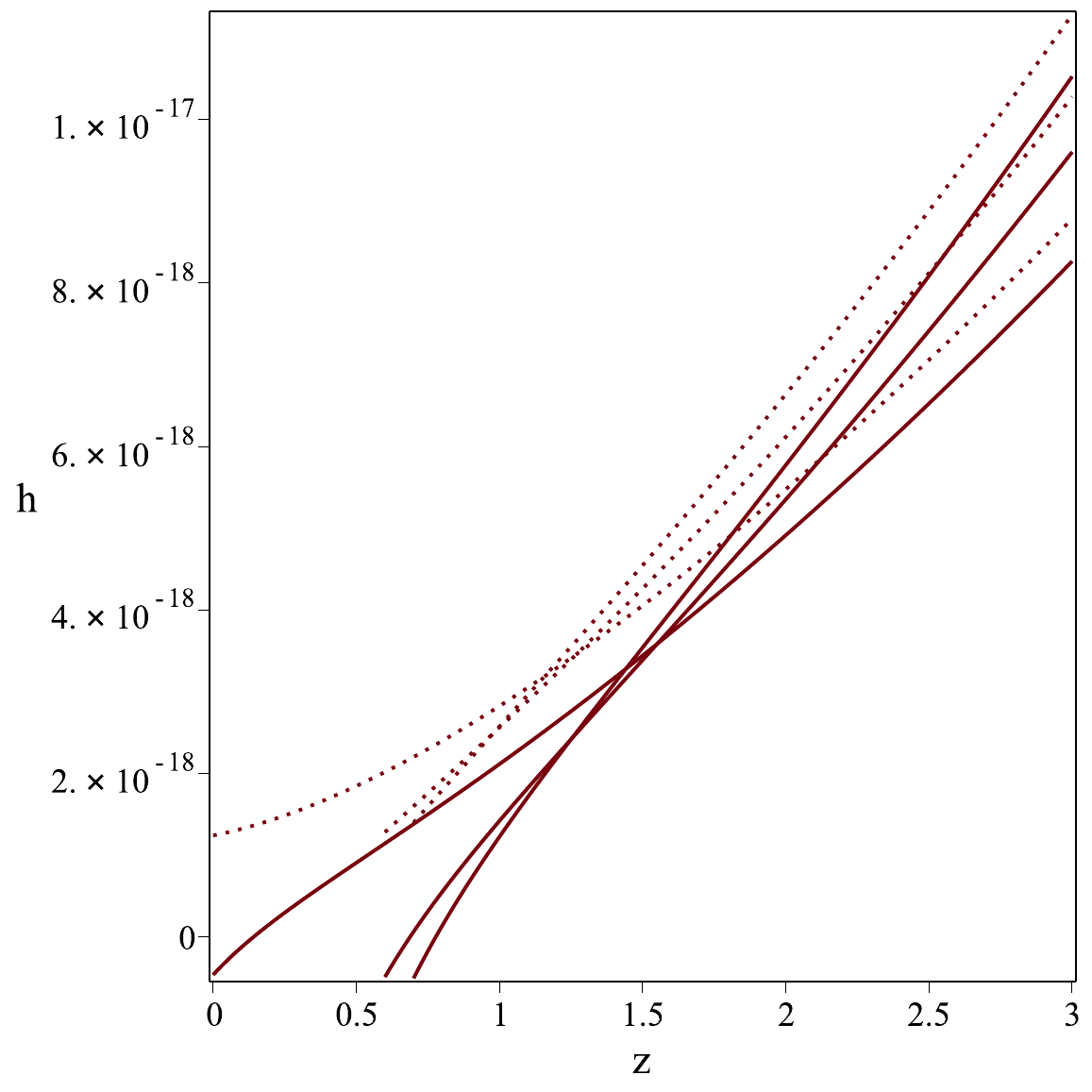}
 \caption{Evolution of $h$ with $z$ for GCG universes. Lines description is as in previous Figs.~\ref{fig:deltas_vs_z1}--\ref{fig:deltas_vs_z3}, \ref{fig:deltas_vs_z5}.}
\label{fig:deltas_vs_z7}
\end{figure*}

\section{Results}
\label{sec:res}

The results of the calculations are plotted in Figs.~\ref{fig:deltas_vs_z1}--\ref{fig:deltas_vs_z7}. We solved the system of 3 differential equations given by Eqs.~(\ref{eq:dot_delta_a}--\ref{eq:dot_theta_a_omega}). 
In the overdensity evolution Eq.~\eqref{eq:dot_delta_a}, $j=1$ corresponds to the baryons overdensity ($\delta_b$), $j=2$ corresponds to the GCG overdensity ($\delta_{\rm GCG}$), while their flow collapse equation is given by Eq.~\eqref{eq:dot_theta_a_omega}. The equations were solved with the same initial conditions used in \citep{DelPopolo2013}, namely $\delta_b(1000)= 10^{-5}$, $\delta_{\rm GCG}(1000)= 3.5 \times 10^{-3}$, and $\theta=0$. 
In Fig.~\ref{fig:deltas_vs_z1}, is plotted the growth of perturbations, $\delta_b$. The solid lines represents $\delta_b$
with $\alpha$ increasing from 0 (bottom solid lines), to 0.5 (central solid lines), and 1 (top solid lines), for $\beta=0$. The dotted lines represent $\delta_b$, with $\beta = 0.01$ (top two panels), $\beta=0.02$ (central two panels), and $\beta=0.04$ (bottom two panels). The value of $\eta_0$ is equal to $4 \times 10^{-3}$ in all left column panels, and 0.03 in all right column panels. As was already noticed by \citep{Fern,DelPopolo2013}, larger values of $\alpha$ produce a faster collapse via larger values of the effective sound speed at lower $z$. The dependence from $\alpha$ in $\delta_b$ comes from the third equation, the evolution of $\theta$, through the effective speed velocity $c^2_{\rm eff}$, while in the case of $\delta_{\rm GCG}$, see Fig.~\ref{fig:deltas_vs_z2} the dependence from $\alpha$ comes from $c^2_{\rm eff}$, and $w$. Moreover, at smaller $z$, when DE dominates, larger values of $\alpha$ produce a later transition from DM to DE dominated stages of the GCG universes. The dashed lines represent how the change of the term $\sigma^2-\omega^2$, parameterized through $\beta$, and $\eta_0$ modifies the collapse. In all the subplots in Fig.~\ref{fig:deltas_vs_z1}, $\delta_b$ changes due to $\alpha$, whose values increase from 0 to 0.5 and finally 1, going from the bottom lines to the top ones. The increase of $\alpha$ accelerates the collapse. In the top left subplot, $\beta=0.01$ for the dotted lines, and $\eta_0=4 \times 10^{-3}$, while in the top right subplot $\beta=0.01$ for the dotted lines, and $\eta_0=0.03$. Both $\beta$, and $\eta_0$ produce a dampening of the growth of the perturbations, but the effect of dynamical friction is larger than that of shear and vorticity. The same trend is shown by the central panels in Fig.~\ref{fig:deltas_vs_z1}, in which we have the same change in $\alpha$, while $\beta=0.02$ for the dotted lines, and $\eta_0=4 \times 10^{-3}$ (left), or $\beta=0.02$ for the dotted lines, and $\eta_0=0.03$ (right). In the bottom subplots, we have the same change in $\alpha$, while $\beta=0.04$ for the dotted lines, and $\eta_0=4 \times 10^{-3}$ (left), or $\beta=0.04$ for the dotted lines, and $\eta_0=0.03$ (right). Summarizing, $\alpha$ accelerates the collapse while $\beta$ and $\eta_0$ dampens it. Apart from the dampening caused by $\beta$ already pointed in \citep{DelPopolo2013}, dynamical friction brings a stronger dampening than that of $\beta$. 
The collapse acceleration produced by larger values of $\alpha$ is mitigated by the additive terms (shear, vorticity, and dynamical friction). Somehow, the effects of the additive terms can be mimicked by a reduction of $\alpha$.


Similarly to Fig.~\ref{fig:deltas_vs_z1}, in Fig.~\ref{fig:deltas_vs_z2}, we show the evolution of $\delta_{\rm GCG}$ when $\alpha$, $\beta$, and $\eta_0$, are modified. The result can be similarly discussed as for $\delta_b$. In Fig.~\ref{fig:deltas_vs_z3}, we plot the evolution of $w_c$, with analogue changes as the previous figures in terms of $\alpha$, $\beta$, and $\eta_0$. As in previous figures, in all subplots, $\alpha$ increase from 0 to 0.5, and finally 1, from  bottom to top lines. The two top subplots are characterized by $\beta=0.01$ for the dotted lines, $\eta_0= 4 \times 10^{-3}$ (left), and $\beta=0.01$ for the dotted lines, and $\eta_0=0.03$ (right). The two central subplots are characterized by $\beta=0.02$ for the dotted lines, $\eta_0= 4 \times 10^{-3}$ (left), and $\beta=0.02$ for the dotted lines, and $\eta_0=0.03$ (right). Finally, the two bottom subplots are characterized by $\beta=0.04$ for the dotted lines, $\eta_0= 4 \times 10^{-3}$ (left), and $\beta=0.04$ for the dotted lines, and $\eta_0=0.03$ (right). Again, $\alpha$ has a strong effect on the results. Larger $\alpha$ produce values of $w_c$ closer to 0 during the collapse and moreover leads to a later transition from DM to DE dominated stages of the GCG universes. Since shear, vorticity, and dynamical friction dampen the collapse, their effect on $w_{\rm c}$ induces a more pronounced departure from zero. 

The quoted result is obtained for a fixed value of $\overline{C}$ ($\overline{C}=0.75$ in our case). If we increase the content of DE of the system, which corresponds to increasing the value of $\overline{C}$, the collapse will happen at later times or it will be prevented, with the occurence that $w_{\rm c}$ will no longer be close to zero.

Fig.~\ref{fig:deltas_vs_z4} represents $w$ given by Eq.~\eqref{eq:w_gcg}, depending on $\overline{C}$, $a$, and $\alpha$ only, and thus independent from shear, vorticity, and dynamical friction, and as a consequence the result is identical to those of \cite{Fern,DelPopolo2013}. The solid line represents the case $\alpha=0$, the  dash-dotted one the case $\alpha=0.5$, and the dashed line the case $\alpha=1$.

In Fig.~\ref{fig:deltas_vs_z5}, we show the evolution of $c^2_{\rm eff}$.  The values of $\alpha$, $\beta$, and $\eta_0$ in the curves in the subplots have the same values as in Figs.~\ref{fig:deltas_vs_z1}--\ref{fig:deltas_vs_z3}. Note that the bottom curves are flat. As the plots show, shear, vorticity, and dynamical friction produce an increase in the value of $c^2_{\rm eff}$. 

Fig.~\ref{fig:deltas_vs_z6}, shows the evolution of $c_s^2$. Since the sound speed is not depending on $\beta$, and $\eta_0$, the result is the same as that in \cite{Fern,DelPopolo2013}. 

A comparison between Fig.~\ref{fig:deltas_vs_z5} and Fig.~\ref{fig:deltas_vs_z6} shows the different behavior of $c^2_{\rm eff}$ and $c_{\rm s}^2$, implying a different behavior of the GCG component locally ($c^2_{\rm eff}$) and in the background ($c_{\rm s}^2$). 

Finally, Fig.~\ref{fig:deltas_vs_z7} shows the evolution of $h=H+\frac{\theta}{3a}$ with $z$. The meaning of the symbols in this figure is the same as those in Figs.~\ref{fig:deltas_vs_z1}--\ref{fig:deltas_vs_z3}, and Fig.~\ref{fig:deltas_vs_z5}.  Larger values of $\alpha$ give rise to a faster decrease in $h$. Since the turn-around redshift, $z_{\rm ta}$, can be defined as the $z$ at which $h=0$, it is clear that higher $\alpha$ imply a larger $z_{\rm ta}$ and an earlier collapse. 

An important point to discuss now, is that previous works \cite[e.g.][]{sandvik:2004,gorini:2008} showed a problem in UDM models, namely oscillations or exponential blowup of the dark matter power spectrum not seen in observations. 
The problem  evident on galactic scales and at recent times, cannot be solved taking baryons into account \cite{Col}. 
Both \cite{sandvik:2004} and \cite{beca}, showed that gravitational effects of DM, at late time, can add fluctuations 
to baryons but that they are unable to erase the ones already present.

Our result, concerning the effect of $\alpha$ on the growth of perturbations, are in agreement with \cite{Fern} and in disagreement with the linear theory of perturbation in GCG universes \cite[e.g.][]{sandvik:2004,gorini:2008}. In our study, the growth dampening of perturbations produced mostly by dynamical friction, and then by shear and rotation,  reduce the possible presence of oscillations as found by \cite{sandvik:2004,gorini:2008}. Moreover, as shown, the effect of dynamical friction increases for smaller scales. In Fig.~\ref{fig:final}, we plot the power spectrum for $\alpha=0$, $10^{-5}$, and $10^{-4}$, with $\beta=0.04$, and $\eta=0.03$. The plot does not show the irregular behavior present in \cite{sandvik:2004}.   

Another important point is that the results are based on a \textit{top-hat} profile for the density, with pressure. A flat profile does not contain pressure gradients and the growth of perturbations can only be suppressed  
by an accelerated expansion. By introducing a non-flat initial perturbation, it would be possible to improve the understanding of how $\alpha$ affects structure formation. 

%
%

%
%

\section{Conclusions}
\label{sec:conc}

In this paper, we studied the perturbations evolution in GCG universes. We extended the \cite{Fern,DelPopolo2013} papers taking not only in consideration shear and vorticity as we did in \citep{DelPopolo2013}, but also dynamical friction. As we already knew from \cite{Fern,DelPopolo2013}, larger values of $\alpha$ speed up the collapse, while shear, vorticity, and dynamical friction  produces a dampening of this acceleration, visible in the figures showing the evolution of $\delta_{\rm b}$, $\delta_{\rm GCG}$, etc. A clear evidence of the difference in the linear and non linear dynamical behavior of the GCG
is shown by the comparison of $w_{\rm c}$, and $c_{\rm eff}^2$ local (non-linear) parameters with $w$, and $c_{\rm s}^2$, global (linear) ones. We found in particular that the role of dynamical friction eliminated the oscillations in the \cite{sandvik:2004}  
spectrum. Notwithstanding, the SCM is a faithful technique to study gravitational collapse and structure formation,   
with results comparable to those of simulations \cite{ascasi}, further improvements of the present paper can be obtained considering a non "top-hat", smooth, profile allowing for spatial pressure gradients. Moreover, more realistic profiles would improve our understanding of the local dynamics of {GCG} universes, and how the background dynamics is influenced by local non-linear inhomogeneities.

\vspace{0.05cm}
\section*{Acknowledgments}
MLeD acknowledges the financial support by the Lanzhou University starting fund, the Fundamental Research Funds for the Central Universities (Grant No. lzujbky-2019-25), National Science Foundation of China  (NSFC grant No.12247101)
and the 111 Project under Grant No. B20063.
\appendix
\section{GCG sound speeds and EoSs}\label{sec:appendixI}

The GCG EoS is given by: 
\begin{equation}
p=-\frac{C}{\rho^{\,\alpha}},
\label{eq:EoS_gCg}
\end{equation}
where $\rho$ is the density, $p$ is the pressure, $C$ and $\alpha$ are positive constants, . 
The standard Chaplygin gas (CG) corresponds to the GCG when $\alpha = 1$. The GCG background density evolution follows
\begin{equation}
 \rho = \rho_0 \left[\bar C + (1-\bar C)a^{-3(\alpha + 1)} \right]^{\frac{1}{1+\alpha}}\;,
\label{eq:density_gCg}
\end{equation}
as in Avelino \Mov{\it et al. }\cite{Ave}, where $a$ is the cosmic scale factor, related to the cosmological redshift as usual by $1 + z = a_0/a$, and $\bar{C}=C/\rho_0^{1+\alpha}$, where $\rho_0$ is the density at the present epoch.
The EoS parameter, $w$, is given by
\begin{equation}
w = -\bar C\left[\bar C + (1-\bar C)a^{-3(\alpha + 1)} \right]^{-1}\;.
\label{eq:w_gcg}
\end{equation}
and $c^2_s=-\alpha w$. Eq.~(\ref{eq:w_gcg}) shows that the GCG behaves as DM at early time ($a \rightarrow 0$),
 and at later times, for $a\gg1$, it follows $w \rightarrow -1$, approaching a DE behavior. The effective sound speed $c^2_{\rm eff}$ employed is the same as that proposed by \cite{Fern}, namely:
\begin{eqnarray}
 c_{\rm eff}^2 =
& = &-\frac{C}{\rho^{1+\alpha}}\frac{(1+\delta)^{-\alpha}-1}{\delta} =
w\frac{(1+\delta)^{-\alpha}-1}{\delta}\;.  \label{eq:my_c2eff} \nonumber \\
\label{eq:cef}
\end{eqnarray}
The effective sound speed, as shown by Eq.~\eqref{eq:cef}, depends on the collapsed region (through $\delta$) and the background (through $w$).The effective $w$ relative to the collapsed region, namely $w_c$, is given by Eq.~(20) of \citep{Fern}
\begin{equation}
 w_c=-\frac{C}{(\rho(1+\delta))^{1+\alpha}}=\frac{w}{(1+\delta)^{1+\alpha}}. \label{eq:wc_exact}
\end{equation}

\section{GCG power spectrum}
In this section, we describe how to derive the power spectrum for the Chaplygin gas by linearizing the perturbation equations and working in Fourier space.

Starting with the continuity equation for a fluid component $j$ in Fourier space, and using the background equations, we obtain the following relation in the linear regime:
\begin{equation}
\delta_j' + \frac{3}{a}(c_{\text{eff}}^2 - w_j)\delta_j = -\frac{(1 + w_j)}{a^2 H} \theta_j,
\end{equation}
where $\theta_j = i\vec{k} \cdot \vec{v}_j$ is the divergence of the velocity field in comoving coordinates. Next, we linearize the Euler equation for the fluid, resulting in:
\begin{equation}
\theta_j' + \left(\frac{1}{a} + \frac{\eta}{H}\right)\theta_j = -\frac{k^2 \phi^2}{a^2 H} - \frac{c_{\text{eff}}^2 k^2 \delta_j}{a^2 H (1 + w_j)} - \frac{\sigma^2 - \omega^2}{a^2 H}.
\end{equation}
To relate the gravitational potential $\phi$ to the density perturbation, we employ the Poisson equation:
\begin{equation}
-\frac{k^2}{a^2} \phi = 4\pi G \sum_k \rho_{0,k} \delta_k (1 + 3c_{\text{eff}}^2).
\end{equation}

Differentiating the continuity equation and using the Euler equation, we obtain the second-order Mukhanov-Sasaki (MS) equation:
\begin{align}
\delta_j'' &+ \left[\frac{1}{a} + \frac{\eta}{H} + \frac{3}{a}(c_{\text{eff},j}^2 - w_j)\right] \delta_j' \nonumber\\
&+ \left[\frac{3}{a^2}(c_{\text{eff},j}^2 - w_j)\left(1 + \frac{\eta}{H}\right)- \frac{3H}{2a^2} \sum_k \Omega_k (1 + 3c_{\text{eff},k}^2)\right] \nonumber\\
&\hspace{0.5cm}\times\delta_j = 0.
\end{align}

In the super-horizon regime ($k \ll aH$), the MS equation leads to the following form:
\begin{align}
\delta_j'' &+ \left[\frac{1}{a} + \frac{\eta}{H} + \frac{3}{a}(c_{\text{eff},j}^2 - w_j)\right] \delta_j' \nonumber\\
&- \frac{3H^2}{2a^2} \Omega_j (1+w_j) (1 + 3c_{\text{eff},j}^2) \delta_j = 0.
\end{align}

Assuming a power-law ansatz for the perturbation, $\delta_j \propto a^\lambda$, we find:
\begin{align}
\delta_j' &= \lambda a^{\lambda - 1},\\
\delta_j'' &= \lambda(\lambda - 1)a^{\lambda - 2}.
\end{align}

Substituting these into the MS equation gives:
\begin{align}
\lambda(\lambda - 1)a^{\lambda - 2} + \left[\frac{1}{a} + \frac{\eta}{H} + \frac{3}{a}(c_{\text{eff},j}^2 - w_j)\right]\lambda a^{\lambda - 1} \nonumber\\
- \frac{3}{2a^2} \Omega_j (1+w_j) (1 + 3c_{\text{eff},j}^2) a^\lambda = 0.
\end{align}

Dividing the entire equation by $a^{\lambda - 2}$ results in:
\begin{align}
\lambda(\lambda - 1) &+ \left[1 + \frac{\eta a}{H} + 3(c_{\text{eff},j}^2 - w_j)\right] \lambda \nonumber \\
&- \frac{3}{2} \Omega_j (1+w_j) (1 + 3c_{\text{eff},j}^2) = 0.
\end{align}

For a Chaplygin gas fluid, $\Omega_j \simeq 1$ and $c_{eff}^2 \ll 1$. Hence the above equation  reduces to:
\begin{equation}
\lambda^2 + (1 - 3w_j)\lambda - \frac{3}{2}(1 + w_j) = 0.
\end{equation}

Solving this quadratic equation for $\lambda$ yields:
\begin{equation}
\lambda = \frac{-(1 - 3w_j) \pm \sqrt{(1 - 3w_j)^2 + 6(1 + w_j)}}{2}.
\end{equation}

For the positive root, $\eta$ modifies the growth rate:
\begin{equation}
\lambda = \frac{3}{2}(1 + w_j) - \frac{\eta}{2H}.
\end{equation}

This shows:
\begin{itemize}
  \item If $\eta > 0$:  growth is slowed.
  \item If $\eta = 0$:  the standard growth scenario is recovered.
\end{itemize}

At horizon crossing ($k = aH$), modes freeze out, and we can evaluate the curvature perturbation $\mathcal{R}$ as:
\begin{equation}
\mathcal{R} = -\frac{5 + 3w_j}{3(1 + w_j)} \delta_j.
\end{equation}
where $|\delta_j|^2$ becomes:
\begin{equation}
|\delta_j|^2 \approx \frac{H^2}{c_{\text{eff},j}^2 k^3}, \hspace{0.5cm}{\rm at}\,\,\, k=aH.
\end{equation}
Hence, the power spectrum of curvature perturbations is:
\begin{align}
P_\mathcal{R}(k)& = \frac{k^3}{2\pi^2} |\mathcal{R}_k|^2 \nonumber\\
&= \left(\frac{5 + 3w}{3(1 + w_j)}\right)^2 \frac{H^2}{2\pi^2 c_{\text{eff},j}^2 } \times \left(1-\frac{\sigma^2-\omega^2}{6H^2}\right).
\end{align}

The spectral index is defined as:
\begin{align}
n_s - 1 &= \frac{d \ln P_\mathcal{R}(k)}{d \ln k}\nonumber \\
& = 3(1 + w_j) - \frac{2\eta}{H} - \frac{d \ln c_{\text{eff},j}^2}{d \ln a},
\end{align}

where the second equality is specific to the Chaplygin gas.

\begin{figure*}[!ht]
\centering
\includegraphics[angle=0,width=0.43\hsize]{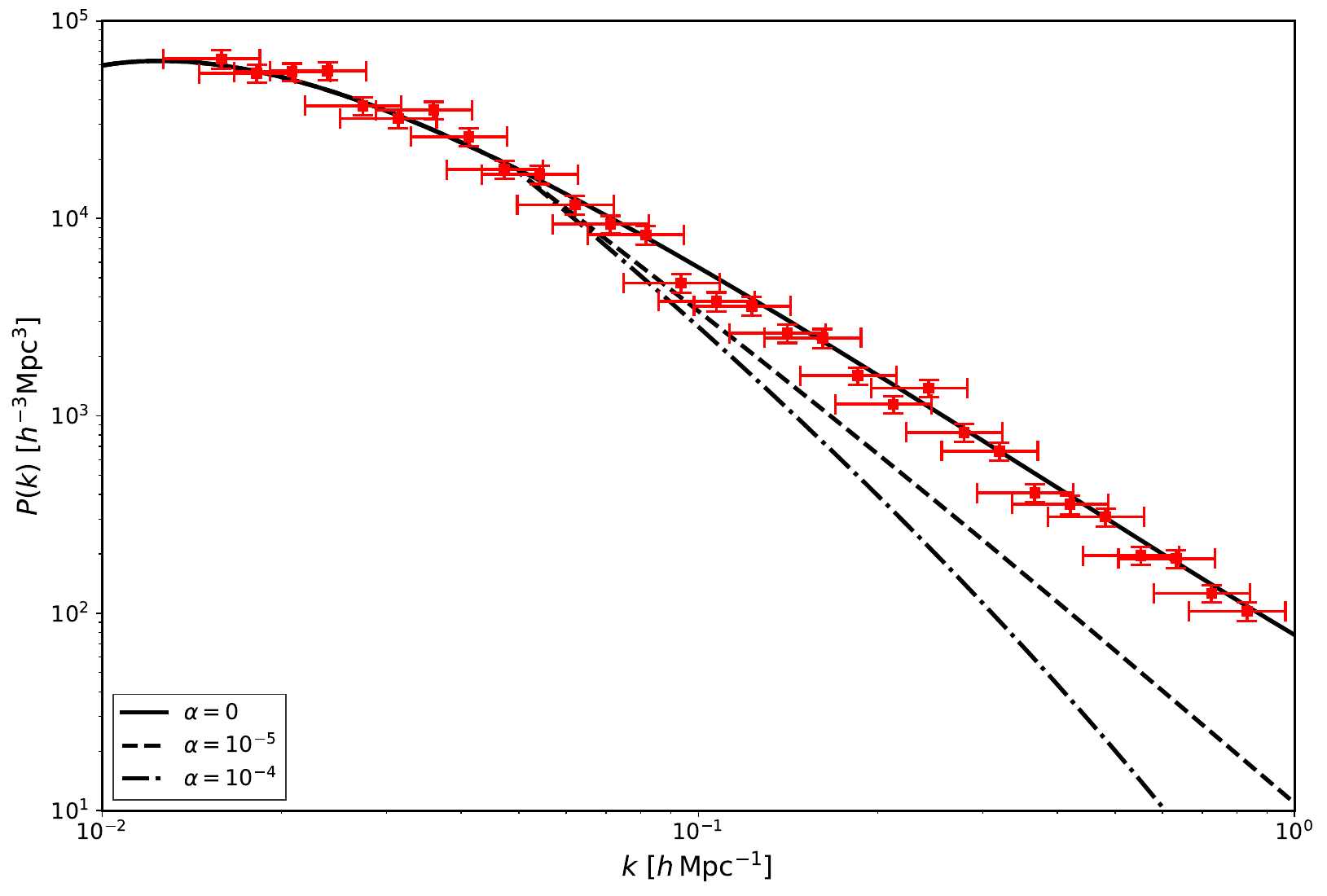}
\caption{The power spectrum of the Chaplygin gas while considering three different values for $\alpha$. The data is related to the 2df Galaxy redshift survey.}
\label{fig:final}
\end{figure*}

Under these considerations, one can obtain the power spectrum of Chaplygin gas as:
\begin{align}
P_\mathcal{R}(k) = \frac{H^2}{2\pi^2 c_{\text{eff},j}^2}\left(\frac{5 + 3w_j}{3(1 + w_j)}\right)^2\!\!\left(1 - \frac{\sigma^2 -\omega^2}{6H^2}\right) \left(\frac{k}{k_0}\right)^{n_s - 1},
\end{align}
where $k_0=0.05$ is the pivot scale. Obviously, this power spectrum includes corrections due to shear and vorticity. In addition, the relations quoted in Appendix~\ref{sec:appendixI} imply the need to set some parameters. In this regard, we
have set $\beta = 0.04$, $n_s = 0.96$, and $\bar{C} = 0.75$. 
Moreover, we have incorporated data for the power spectrum of the 2df galaxy redshift survey in Fig. \ref{fig:final}, where we have shown the power spectrum of the Chaplygin gas for three values of $\alpha = 0, 10^{-5}$ and $10^{-4}$.

\end{document}